\documentclass[english]{revtex4}
\usepackage[T1]{fontenc}
\usepackage[latin9]{inputenc}
\usepackage{color,times}
\usepackage{amsmath}
\usepackage{amssymb}
\usepackage{graphicx}
\usepackage{epstopdf}
\usepackage{epsfig}
\usepackage{esint}

\makeatletter
\@ifundefined{textcolor}{}
{%
 \definecolor{BLACK}{gray}{0}
 \definecolor{WHITE}{gray}{1}
 \definecolor{RED}{rgb}{1,0,0}
 \definecolor{GREEN}{rgb}{0,1,0}
 \definecolor{BLUE}{rgb}{0,0,1}
 \definecolor{CYAN}{cmyk}{1,0,0,0}
 \definecolor{MAGENTA}{cmyk}{0,1,0,0}
 \definecolor{YELLOW}{cmyk}{0,0,1,0}
}

\usepackage{babel}

\usepackage{babel}

\makeatother

\usepackage{babel}
\begin{document}

\title{Quasiprobability distributions in open quantum systems: spin-qubit
systems}

\author{Kishore Thapliyal$^{a}$}

\author{Subhashish Banerjee$^{b}$}

\author{Anirban Pathak$^{a}$}

\email{anirban.pathak@gmail.com, Phone: +91 9717066494}

\author{S. Omkar$^{c}$}

\author{V. Ravishankar$^{d}$ }

\affiliation{$^{a}$Jaypee Institute of Information Technology, A-10, Sector-62,
Noida, UP-201307, India\\
 $^{b}$Indian Institute of Technology Jodhpur, Jodhpur 342011, India\\
 $^{c}$Poornaprajna Institute of Scientific Research, Bengaluru,
India\\
 $^{d}$Indian Institute of Technology Delhi Hauz Khas, New Delhi-110016,
India}

\date{\today}
\begin{abstract}
We study nonclassical features in a number of spin-qubit systems including
single, two and three qubit states, as well as an $N$ qubit Dicke
model and a spin-1 system, of importance in the fields of quantum
optics and information. This is done by analyzing the behavior of
the well known Wigner, $P$, and $Q$ quasiprobability distributions
on them. We also discuss the not so well known $F$ function and specify
its relation to the Wigner function. Here we provide a comprehensive
analysis of quasiprobability distributions for spin-qubit systems
under general open system effects, including both pure dephasing as
well as dissipation. This makes it relevant from the perspective of
experimental implementation. 
\end{abstract}
\maketitle

\section{{\normalsize{{Introduction}}}}

A very useful concept in the analysis of the dynamics of classical
systems is the notion of phase space. A straightforward extension
of this to the realm of quantum mechanics is however foiled due to
the uncertainty principle. Despite this, it is possible to construct
quasiprobability distributions (QDs) for quantum mechanical systems
in analogy with their classical counterparts \citep{KS,SZ,Sch,GSA,Puri,Klimnov}.
These QDs are very useful in that they provide a quantum classical
correspondence and facilitate the calculation of quantum mechanical
averages in close analogy to classical phase space averages. Nevertheless,
the QDs are not probability distributions as they can take negative
values as well, a feature that could be used for the identification
of quantumness in a system.

The first such QD was developed by Wigner resulting in the epithet
Wigner function ($W$) \citep{wig,moyal,hillery,kimnoz,adam1,adam2}.
Another, very well known, QD is the $P$ function whose development
was a precursor to the evolution of the field of quantum optics. This
was originally developed from the possibility of expressing any state
of the radiation field in terms of a diagonal sum over coherent states
\citep{glaub,sudar}. The $P$ function can become singular for quantum
states, a feature that promoted the development of other QDs such
as the $Q$ function \citep{mehta,kano,husimi} as well as further
highlighted the use of the $W$ function which does not have this
feature. These QDs are intimately related to the problem of operator
orderings. Thus, the $P$ and $Q$ functions are related to the normal
and antinormal orderings, respectively, while the $W$ function is
associated with symmetric operator ordering. It is quite clear that
there can be other QDs, apart from the above three, depending upon
the operator ordering. However, among all the possible QDs the above
three QDs are the most widely studied. There exist several reasons
behind the intense interest in these QDs. They can be used to identify
the nonclassical (quantum) nature of a state \cite{referee}. Specifically, nonpositive
values of $P$ function define a nonclassical state. Nonpositivity
of $P$ is a necessary and sufficient criterion for nonclassicality,
but other QDs provide only sufficient criteria.

A nonclassical state can be used to perform tasks that are classically
impossible. This fact motivated many studies on nonclassical states,
for example, studies on squeezed, antibunched and entangled states.
The interest on nonclassical states has been considerably amplified
in the recent past after the advent of quantum information where several
applications of nonclassical states, in particular, of entangled states,
have been reported \citep{my book}. Interestingly many of these applications
have been designed using spin-qubit systems.

Quantum optics deals with atom-field interactions. The atoms, in their
simplest forms, are modeled as qubits (two-level systems). These are
also of immense practical importance as they can be the effective
realizations of Rydberg atoms \citep{sa,ga}. Atomic systems are also
studied in the context of the Dicke model \citep{dicke,dowling},
a collection of two-level atoms; in atomic traps \citep{we}, atomic
interferometers \citep{kita}, polarization optics \citep{kara},
and have recently found applications in quantum computation (\citep{Qu-comp1,Qu-comp2,Qu-comp3,Qu-comp4,Qu-comp5,Qu-comp6}
and references therein) as well as in the generation of long-distance
entanglement \citep{Qu-comp7}. All these would evoke the question
whether one could have QDs for such atomic systems as well. Such questions,
which are of relevance to the present work, would be closely tied
to the problem of development of QDs for $SU(2)$, spin-like (spin-$j$),
systems. Such a development was made in \citep{strato}, where a QD
on the sphere, naturally related to the $SU(2)$ dynamical group \citep{klim,chuma},
was obtained. There are by now a number of constructions of spin QDs
\citep{wootters,vourdas,chatur,leon,paz}, among others.

However, another approach, the one adapted here, is to make use of
the connection of $SU(2)$ geometry to that of a sphere. The spherical
harmonics provide a natural basis for functions on the sphere. This,
along with the general theory of multipole operators \citep{blum,zare},
can be made use of to construct QDs of spin (qubit) systems as functions
of polar and azimuthal angles \citep{agarwal}. Other constructions,
in the literature, of $W$ functions for spin-$1/2$ systems can be
found in \citep{cohen,varilly}, among others. A concept that played
an important role in the above developments, was the atomic coherent
state \citep{arecchi}, which lead to the definition of atomic $P$
function in close analogy to their radiation field counterparts. Another
related development, following \citep{mh} where joint probability
distributions were obtained for spin-1 systems exposed to quadrupole
fields, was a QD obtained from the Fourier inversion of the characteristic
function of the corresponding probability mass function, using the
Wigner-Weyl correspondence. This could be called the characteristic
function or $F$-function approach \citep{rama}.

The fields of quantum optics and information have matured to the point
where intense experimental investigations are being made. Both from
the fundamental perspective as well as from the viewpoint of practical
realizations, it is imperative to study the evolution of the system
of interest taking into account the effect of its ambient environment.
This is achieved systematically by using the formalism of Open Quantum
Systems \citep{louis,bp,pi,sbqbm}.

In the present work, we investigate nonclassicality in a number of
spin-qubit systems including single, two and three qubit states, as
well as $N$ qubit Dicke states and a spin-1 system, of importance
in the fields of quantum optics and information. This is done by analyzing
the behavior of the well known $W$, $P$, $Q$ QDs on them. The significance
of this is rooted to the phenomena of quantum state engineering, which
involves the generation and manipulation of nonclassical states \citep{adam-SE,SE-nature}.
In this context, it is imperative to have an understanding over quantum
to classical transitions, under ambient conditions. Such an understanding
is made possible by the present work, where investigations are done
in the presence of open system effects, both purely dephasing (decoherence)
\citep{sbsrikgp,QND}, also known as quantum non-demolition (QND),
as well as dissipation \citep{sbsrikgp,SGAD}. These aspects of open
system evolution have been realized in a series of beautiful experiments
\citep{haroche,turchette}. We also discuss the not so well known
$F$ function and specify its relation to the $W$ function. Further,
we expect this work to have an impact on tomography related issues,
as borne out in \citep{agarwal98}, where a method for quantum state
reconstruction of a system of spins or qubits was proposed using the
$Q$ function. Also, the $Q$ function, studied here, can be turned
to address fundamental issues such as complementarity between number
and phase distributions \citep{shapiro,hall,agarwalphase}, under
the influence of QND as well as dissipative interactions with their
environment, as well as for phase dispersion in atomic systems \citep{sbphase,sbsrik}.
Here, to the best of our knowledge, we provide, for the first time,
a comprehensive analysis of QDs for spin-qubit systems under general
open system effects.

The plan of this paper is as follows. In the next section, we will
briefly discuss the QDs that will be subsequently used in the rest
of the work, i.e., the $W$, $P$, $Q$, and $F$ functions. This
will be followed by a study of open system QDs for single qubit states.
Next, we take up the case of some interesting two and three qubit
states as well as the well known $N$ qubit Dicke model. We then discuss,
briefly, QDs of a spin-1 system. These examples will provide an understanding
of quantum to classical transitions as indicated by the various QDs,
under general open system evolutions. Although QDs have been frequently
used to identify the existence of nonclassical states \citep{perina-book},
they do not directly provide any quantitative measure of the amount
of nonclassicality. Keeping these in mind, several measures of nonclassicality
have been proposed, but all of them are seen to suffer from some limitations
\citep{with-adam-meas}. A specific measure of nonclassicality is
the nonclassical volume, which considers the doubled volume of the integrated negative part of
the $W$ function as a measure of nonclassicality \citep{zyco}. 
 In the penultimate section, we make a study of quantumness, in some
of the systems considered in this work, by using nonclassical volume
\citep{zyco}. We then make our conclusions.

\section{{\normalsize{{{Distribution functions for spin (qubit) systems}}}}}

Here, we briefly discuss the different QDs, i.e., the $W$, $P$,
$Q$, and $F$ functions, subsequently used in the paper.

\subsection{The Wigner function}

Exploiting the connection between spin-like, $SU(2)$, systems and
the sphere, a QD can be expressed as a function of the polar and azimuthal
angles. This expanded over a complete basis set, a convenient one
being the spherical harmonics, the $W$ function for a single spin-$j$
state can be expressed as \citep{agarwal} 
\begin{equation}
\begin{array}{lcl}
W\left(\theta,\phi\right) & = & \left(\frac{2j+1}{4\pi}\right)^{1/2}\underset{K,Q}{\sum}\rho_{KQ}Y_{KQ}\left(\theta,\phi\right),\end{array}\label{eq:wigner-singlequbit}
\end{equation}
where $K=0,1,\ldots,2j$, and $Q=-K,-K+1,\ldots,0,\ldots,K-1,K$,
and 
\begin{equation}
\begin{array}{lcl}
\rho_{KQ} & = & Tr\left\{ T_{KQ}^{\dagger}\rho\right\} .\end{array}\label{eq:rho_kq}
\end{equation}
Here, $Y_{KQ}$ are spherical harmonics and $T_{KQ}$ are multipole
operators given by 
\begin{equation}
\begin{array}{lcl}
T_{KQ} & = & \underset{m,m^{\prime}}{\sum}\left(-1\right)^{j-m}\left(2K+1\right)^{1/2}\left(\begin{array}{ccc}
j & K & j\\
-m & Q & m^{\prime}
\end{array}\right)|j,m\rangle\langle j,m^{\prime}|,\end{array}\label{eq:multipole-operator}
\end{equation}
where $\left(\begin{array}{ccc}
j_{1} & j_{2} & j\\
m_{1} & m_{2} & m
\end{array}\right)=\frac{\left(-1\right)^{j_{1}-j_{2}-m}}{\sqrt{2j+1}}\langle j_{1}m_{1}j_{2}m_{2}|j-m\rangle$ is the Wigner $3j$ symbol \citep{varshalo} and $\langle j_{1}m_{1}j_{2}m_{2}|j-m\rangle$
is the Clebsh-Gordon coefficient. The multipole operators $T_{KQ}$
are orthogonal to each other and they form a complete set with property
$T_{KQ}^{\dagger}=\left(-1\right)^{Q}T_{K,-Q}$. The $W$ function
is normalized as 
\[
\int W\left(\theta,\phi\right)\sin\theta d\theta d\phi=1,
\]
and $W^{*}\left(\theta,\phi\right)=W\left(\theta,\phi\right)$. Similarly,
the $W$ function of a two particle system, each with spin-$j$ is \citep{agarwal,rama}
\begin{equation}
\begin{array}{lcl}
W\left(\theta_{1},\phi_{1},\theta_{2},\phi_{2}\right) & = & \left(\frac{2j+1}{4\pi}\right)\underset{K_{1},Q_{1}}{\sum}\underset{K_{2},Q_{2}}{\sum}\rho_{K_{1}Q_{1}K_{2}Q_{2}}Y_{K_{1}Q_{1}}\left(\theta_{1},\phi_{1}\right)Y_{K_{2}Q_{2}}\left(\theta_{2},\phi_{2}\right),\end{array}\label{eq:wigner2qubit}
\end{equation}
where $\begin{array}{lcl}
\rho_{K_{1}Q_{1}K_{2}Q_{2}} & = & Tr\left\{ \rho T_{K_{1}Q_{1}}^{\dagger}T_{K_{2}Q_{2}}^{\dagger}\right\} .\end{array}$ Here, $W\left(\theta_{1},\phi_{1},\theta_{2},\phi_{2}\right)$ is also
normalized as 
\[
\int W\left(\theta_{1},\phi_{1},\theta_{2},\phi_{2}\right)\sin\theta_{1}\sin\theta_{2}d\theta_{1}d\phi_{1}d\theta_{2}d\phi_{2}=1.
\]
Further, it is known that any arbitrary operator can be mapped into
the $W$ function or any other QD discussed here. In what follows,
using the same notations we describe $P$, $Q$ and $F$ functions
for single spin-$j$ state and for two spin-$j$ particles. It may
be noted that all the analytic expressions for the QDs given below
are normalized.

\subsection{The $P$ function}

In analogy with the $P$ function for continuous variable systems, the $P$ function 
for a single spin-$j$ state is defined as \cite{agarwal}
\begin{equation}
\rho = \int d\theta  d\phi P\left(\theta, \phi\right) |\theta, \phi \rangle \langle \theta, \phi|,
\end{equation}
and can be shown to be 
\begin{equation}
\begin{array}{lcl}
P\left(\theta,\phi\right) & = & 
\underset{K,Q}{\sum}\rho_{KQ}Y_{KQ}\left(\theta,\phi\right)\left(\frac{1}{4\pi}\right)^{1/2}
\left(-1\right)^{K-Q}\left(\frac{\left(2j-K\right)!\left(2j+K+1\right)!}{\left(2j\right)!\left(2j\right)!}
\right)^{1/2}.\end{array}\label{eq:Pfunction-singlequbit}
\end{equation}
The $P$ function for two spin-$j$ particles is \citep{agarwal,rama}
\begin{equation}
\begin{array}{lcl}
P\left(\theta_{1},\phi_{1},\theta_{2},\phi_{2}\right) & = & \underset{K_{1},Q_{1}}{\sum}\underset{K_{2},Q_{2}}{\sum}\rho_{K_{1}Q_{1}K_{2}Q_{2}}Y_{K_{1}Q_{1}}\left(\theta_{1},\phi_{1}\right)Y_{K_{2}Q_{2}}\left(\theta_{2},\phi_{2}\right)\\
 & \times & \left(-1\right)^{K_{1}-Q_{1}+K_{2}-Q_{2}}\left(\frac{1}{4\pi}\right)\left(\frac{\sqrt{\left(2j-K_{1}\right)!\left(2j-K_{2}\right)!\left(2j+K_{1}+1\right)!\left(2j+K_{2}+1\right)!}}{\left(2j\right)!\left(2j\right)!}\right).
\end{array}\label{eq:Pfunction-2qubit}
\end{equation}
Here $|\theta, \phi \rangle$ is the atomic coherent state \citep{arecchi} and can be expressed in terms of
the Wigner-Dicke states $|j, m \rangle$ as
\begin{equation}
|\theta, \phi \rangle = \underset{m=-j}{\sum^{j}}  \left(\begin{array}{c}
2j \\
m+j 
\end{array}\right)^{1/2}\sin^{j+m}(\frac{\theta}{2}) \cos^{j-m}(\frac{\theta}{2}) e^{-i(j+m)\phi} |j,m \rangle. \label{eq:atomic-coherent-state}
\end{equation}

\subsection{The $Q$ function}

Similarly, the $Q$ function for a single spin-$j$ state is  
\begin{equation}
Q\left(\theta,\phi\right) = \frac{2j+1}{4\pi}  \langle \theta, \phi| \rho | \theta, \phi \rangle,
\end{equation}
and can be expressed as \cite{agarwal}
\begin{equation}
\begin{array}{lcl}
Q\left(\theta,\phi\right) & = & \underset{K,Q}{\sum}\rho_{KQ}Y_{KQ}\left(\theta,\phi\right)\left(\frac{1}{4\pi}\right)^{1/2}\left(-1\right)^{K-Q}\left(2j+1\right)\left(\frac{\left(2j\right)!\left(2j\right)!}{\left(2j-K\right)!\left(2j+K+1\right)!}\right)^{1/2}.\end{array}\label{eq:Qfunction-singlequbit}
\end{equation}

Further, the normalized $Q$ function for two particle system of spin-$j$ \citep{agarwal,rama}
particles is 
\begin{equation}
\begin{array}{lcl}
Q\left(\theta_{1},\phi_{1},\theta_{2},\phi_{2}\right) & = & \underset{K_{1},Q_{1}}{\sum}\underset{K_{2},Q_{2}}{\sum}\rho_{K_{1}Q_{1}K_{2}Q_{2}}Y_{K_{1}Q_{1}}\left(\theta_{1},\phi_{1}\right)Y_{K_{2}Q_{2}}\left(\theta_{2},\phi_{2}\right)\left(\frac{\left(2j+1\right)^{2}}{4\pi}\right)\\
 & \times & \left(-1\right)^{K_{1}-Q_{1}+K_{2}-Q_{2}}\left(\frac{\left(2j\right)!\left(2j\right)!}{\sqrt{\left(2j-K_{1}\right)!\left(2j-K_{2}\right)!\left(2j+K_{1}+1\right)!\left(2j+K_{2}+1\right)!}}\right).
\end{array}\label{eq:Qfunction-2qubit}
\end{equation}

\subsection{The $F$ function}

The $F$ distribution function \citep{rama} is defined using the
relation between Fano statistical tensors and state multipole operators.
Specifically, for a single spin-$j$ state, it is defined as \citep{rama}
\begin{equation}
\begin{array}{lcl}
F\left(\theta,\phi\right) & = & \underset{K,Q}{\sum}\rho_{KQ}Y_{KQ}\left(\theta,\phi\right)\left(\frac{1}{4\pi}\right)^{1/2}\frac{1}{2^{K}}\left(\frac{\left(2j+K+1\right)!}{\left(2j-K\right)!\left\{ j\left(j+1\right)\right\} ^{K}}\right)^{1/2}.\end{array}\label{eq:Ffunction-singlequbit}
\end{equation}
Similarly, the normalized $F$ function for a two particle, spin-$j$, \citep{rama}
system is 
\begin{equation}
\begin{array}{lcl}
F\left(\theta_{1},\phi_{1},\theta_{2},\phi_{2}\right) & = & \underset{K_{1},Q_{1}}{\sum}\underset{K_{2},Q_{2}}{\sum}\rho_{K_{1}Q_{1}K_{2}Q_{2}}Y_{K_{1}Q_{1}}\left(\theta_{1},\phi_{1}\right)Y_{K_{2}Q_{2}}\left(\theta_{2},\phi_{2}\right)\\
 & \times & \left(\frac{1}{4\pi\left(2^{K_{1}+K_{2}}\right)}\right)\left(\frac{\left(2j+K_{1}+1\right)!\left(2j+K_{2}+1\right)!}{\left(2j-K_{1}\right)!\left(2j-K_{2}\right)!\left\{ j\left(j+1\right)\right\} ^{K_{1}+K_{2}}}\right)^{1/2}.
\end{array}\label{eq:Ffunction-2qubit}
\end{equation}

To summarize, all the QDs discussed in this work are normalized to
unity. They are also real functions as they correspond to probability
density functions for classical states. The density matrix of a quantum
state can be reconstructed from these QDs \citep{Klimnov}. One can
also calculate the expectation value of an operator from them \citep{agarwal}.

It would be appropriate here to make a brief comparison of the QDs, discussed above, with their continuous
variable counterparts. The coherent state and thereby the displacement operator $D$, which generate coherent states 
from vacuum and is usually expressed as
$D(\phi) = e^{\phi \hat{a}^{\dagger} - \phi^* \hat{a}}$ with $\hat{a}$, $\hat{a}^{\dagger}$ being the annihilation and 
creation operators of the given Fock space, respectively, play a central role in these considerations. 
Thus, for example, the Wigner function, 
associated with a state $\rho$ is the symplectic Fourier transform of the mean value of $D$ in the state $\rho$, leading to the
standard representation of the Wigner function as the Fourier transform of the skewed matrix representation of $\rho$, 
\begin{equation}
 W(\eta, \overline{\eta}) = \frac{1}{2 \pi}\int_{-\infty}^{\infty} d\xi \langle a - \frac{\xi}{2}|\rho | a + 
 \frac{\xi}{2} \rangle e^{i \xi b}, \label{Wusual}
\end{equation}
with $a$ and $b$ being real and $\eta = \frac{1}{\sqrt{2}} (a + i b)$. Similarly, the $P$ function is associated with the
diagonal representation of the state $\rho$ in terms of the coherent state, while the $Q$ function is related to the 
expectation value of $\rho$, with respect to  coherent states. 

The multipole operators $T_{KQ}$ (\ref{eq:multipole-operator})
play a pivotal role in the construction of QDs of spin systems, discussed here. These operators are extensively used 
in the study of atomic and nuclear radiation and can be shown to have properties analogous to those of the coherent
state dispalcement operator $D$ for usual continuous variable bosonic systems \citep{agarwal}. In this sense, the properties
of spin QDs are analogous to those of their continuous variable counterparts, with the atomic coherent state playing the
role of the usual coherent state. Thus, for example, even though, for the spin (qubit) systems, both $P$ and
$W$ QDs are witnesses of quantum correlations, in the sense that their negative values indicate quantumnes in the system,
it is possible for a scenario wherein the $P$ function is negative and $W$ is positive, but not viceversa. The $Q$ function
is always positive, while the $F$ function is same as the $W$ function for spin-$\frac{1}{2}$ systems, as shown below.

Before proceeding further, it is worth noting here that for all the
spin-$\frac{1}{2}$ states (qubits), single or multi-qubit, the $W$
and $F$ QDs are identical. Specifically, for the single qubit case,
the $W$ function is 
\[
W_{\frac{1}{2}}\left(\theta,\phi\right)=\frac{1}{\sqrt{2\pi}}\underset{K,Q}{\sum}M_{K,Q}\left(\theta,\phi\right),
\]
where $M_{K,Q}\left(\theta,\phi\right)=\rho_{KQ}Y_{KQ}\left(\theta,\phi\right)$,
while, the $F$ function is 
\[
F_{\frac{1}{2}}\left(\theta,\phi\right)=\frac{1}{\sqrt{4\pi}}\underset{K,Q}{\sum}M_{K,Q}\left(\theta,\phi\right)\left(\frac{\left(2+K\right)!}{3^{K}\left(1-K\right)!}\right)^{1/2},
\]
where the term inside the brackets with square root is $2$ for both
the values of $K$ (i.e., $0$ or $1$). Similarly, for two spin-$\frac{1}{2}$
states, the $W$ and $F$ functions are 
\[
\begin{array}{lcl}
F_{\frac{1}{2},\frac{1}{2}}\left(\theta_{1},\phi_{1},\theta_{2},\phi_{2}\right) & = & \frac{1}{4\pi}\underset{K_{1},Q_{1}}{\sum}\underset{K_{2},Q_{2}}{\sum}M_{K_{1}Q_{1}K_{2}Q_{2}}\left(\theta_{1},\phi_{1},\theta_{2},\phi_{2}\right)\left(\frac{\left(2+K_{1}\right)!\left(2+K_{2}\right)!}{3^{K_{1}+K_{2}}\left(1-K_{1}\right)!\left(1-K_{2}\right)!}\right)^{1/2}\\
 & = & \frac{1}{2\pi}\underset{K_{1},Q_{1}}{\sum}\underset{K_{2},Q_{2}}{\sum}M_{K_{1}Q_{1}K_{2}Q_{2}}\left(\theta_{1},\phi_{1},\theta_{2},\phi_{2}\right)\\
 & = & W_{\frac{1}{2},\frac{1}{2}}\left(\theta_{1},\phi_{1},\theta_{2},\phi_{2}\right),
\end{array}
\]
where $M_{K_{1}Q_{1}K_{2}Q_{2}}\left(\theta_{1},\phi_{1},\theta_{2},\phi_{2}\right)=\rho_{K_{1}Q_{1}K_{2}Q_{2}}Y_{K_{1}Q_{1}}\left(\theta_{1},\phi_{1}\right)Y_{K_{2}Q_{2}}\left(\theta_{2},\phi_{2}\right),$
and the term in the brackets is $4$ for all the possible values of
$K_{1}$ and $K_{2}$. This can further be extended for higher number
of spin-$\frac{1}{2}$ states.

Since the $W$ and $F$ functions are the same for spin-$\frac{1}{2}$
systems, we will not discuss the evolution of the $F$ function of
these systems.

\section{{\normalsize{{Distribution functions for single spin-$\frac{1}{2}$
states}}}}

Here, we consider single spin-$\frac{1}{2}$ states, initially in
an atomic coherent state, in the presence of two different noises,
i.e., QND \citep{sbsrikgp,QND}, which are purely dephasing, and the
dissipative SGAD (Squeezed Generalized Amplitude Damping) \citep{sbsrikgp,SGAD}
noises. For calculating the QDs, we will require multipole operators
for $j=\frac{1}{2}$ and $m,\, m^{\prime}=\pm\frac{1}{2}$, giving
$K=0$ and $1$. For $K=0$, $Q=0$, and for $K=1$, $Q=1,\,0,\,-1$.
Using these, the multipole operators $T_{KQ}$ can be obtained as
$\begin{array}{lcl}
T_{00} & = & \frac{1}{\sqrt{2}}\left[\begin{array}{cc}
1 & 0\\
0 & 1
\end{array}\right],\end{array}$ $\begin{array}{lcl}
T_{11} & = & \left[\begin{array}{cc}
0 & 0\\
-1 & 0
\end{array}\right],\end{array}$ $\begin{array}{lcl}
T_{10} & = & \frac{1}{\sqrt{2}}\left[\begin{array}{cc}
1 & 0\\
0 & -1
\end{array}\right],\end{array}$and $\begin{array}{lcl}
T_{1-1} & = & \left[\begin{array}{cc}
0 & 1\\
0 & 0
\end{array}\right].\end{array}$

\subsection{Atomic coherent state in QND noise}

The master equation of the system interacting with a squeezed thermal
bath and undergoing a QND evolution \citep{QND} is 
\begin{equation}
\begin{array}{lcl}
\dot{\rho}{}_{nm}\left(t\right) & = & \left[-\frac{i}{\hbar}\left(E_{n}-E_{m}\right)+i\dot{\eta}\left(t\right)\left(E_{n}^{2}-E_{m}^{2}\right)-\left(E_{n}-E_{m}\right)^{2}\dot{\gamma}\left(t\right)\right]\rho_{nm}\left(t\right),\end{array}\label{eq:master-eq-QND}
\end{equation}
where $E_{n}$s are the eigenvalues of the system Hamiltonian in the
system eigenbasis $|n\rangle$, which here would correspond to the
Wigner-Dicke states \citep{GSA}; 
\[
\eta\left(t\right)=-\underset{k}{\sum}\frac{g_{k}^{2}}{\hbar^{2}\omega_{k}^{2}}\sin\left(\omega_{k}t\right)
\]
and 
\[
\begin{array}{lcl}
\gamma\left(t\right) & = & \frac{1}{2}\underset{k}{\sum}\frac{g_{k}^{2}}{\hbar^{2}\omega_{k}^{2}}\coth\left(\frac{\beta\hbar\omega_{k}}{2}\right)\left|\left(e^{i\omega_{k}t}-1\right)\cosh\left(r_{k}\right)+\left(e^{-i\omega_{k}t}-1\right)\sinh\left(r_{k}\right)e^{2i\Phi_{k}}\right|^{2}.\end{array}
\]
Here, $\beta=\frac{1}{k_{B}T}$, and $k_{B}$ is the Boltzmann constant,
while $r_{k}$ and $\Phi_{k}$ are the squeezing parameters. The initial
density matrix for the atomic coherent state is 
\begin{equation}
\rho\left(0\right)=|\alpha,\beta\rangle\langle\alpha,\beta|,\label{eq:initial-state-QND}
\end{equation}
where $|\alpha,\beta\rangle$ is given by Eq. (\ref{eq:atomic-coherent-state}).
The density matrix (\ref{eq:initial-state-QND}) in the presence of
QND noise at time $t$ becomes 
\begin{equation}
\begin{array}{lcl}
\rho_{jm,jn}\left(t\right) & = & e^{-i\omega\left(m-n\right)t}e^{i\left(\hbar\omega\right)^{2}\left(m^{2}-n^{2}\right)\eta\left(t\right)}e^{-\left(\hbar\omega\right)^{2}\left(m-n\right)^{2}\gamma\left(t\right)}\rho_{jm,jn}\left(0\right),\end{array}\label{eq:QND-densitymatirix}
\end{equation}
where 
\begin{equation}
\begin{array}{lcl}
\rho_{jm,jn}\left(0\right) & = & \langle j,m|\rho\left(0\right)|j,n\rangle\\
 & = & \langle j,m|\alpha,\beta\rangle\langle\alpha,\beta|j,n\rangle.
\end{array}\label{eq:at_t0}
\end{equation}
For $j=\frac{1}{2}$, the initial density matrix is 
\begin{equation}
\rho\left(0\right)=\left[\begin{array}{cc}
\sin^{2}\left(\frac{\alpha}{2}\right) & \frac{1}{2}e^{-i\beta}\sin\alpha\\
\frac{1}{2}e^{i\beta}\sin\alpha & \cos^{2}\left(\frac{\alpha}{2}\right)
\end{array}\right],\label{eq:density-matrix}
\end{equation}
which in the presence of QND noise becomes 
\begin{equation}
\rho\left(t\right)=\left[\begin{array}{cc}
\sin^{2}\left(\frac{\alpha}{2}\right) & \frac{1}{2}e^{-i\omega t}e^{-\left(\hbar\omega\right)^{2}\gamma\left(t\right)}e^{-i\beta}\sin\alpha\\
\frac{1}{2}e^{i\omega t}e^{-\left(\hbar\omega\right)^{2}\gamma\left(t\right)}e^{i\beta}\sin\alpha & \cos^{2}\left(\frac{\alpha}{2}\right)
\end{array}\right].\label{eq:density-matrix-QND}
\end{equation}

\begin{figure}
\begin{centering}
\includegraphics[angle=-90,scale=0.65]{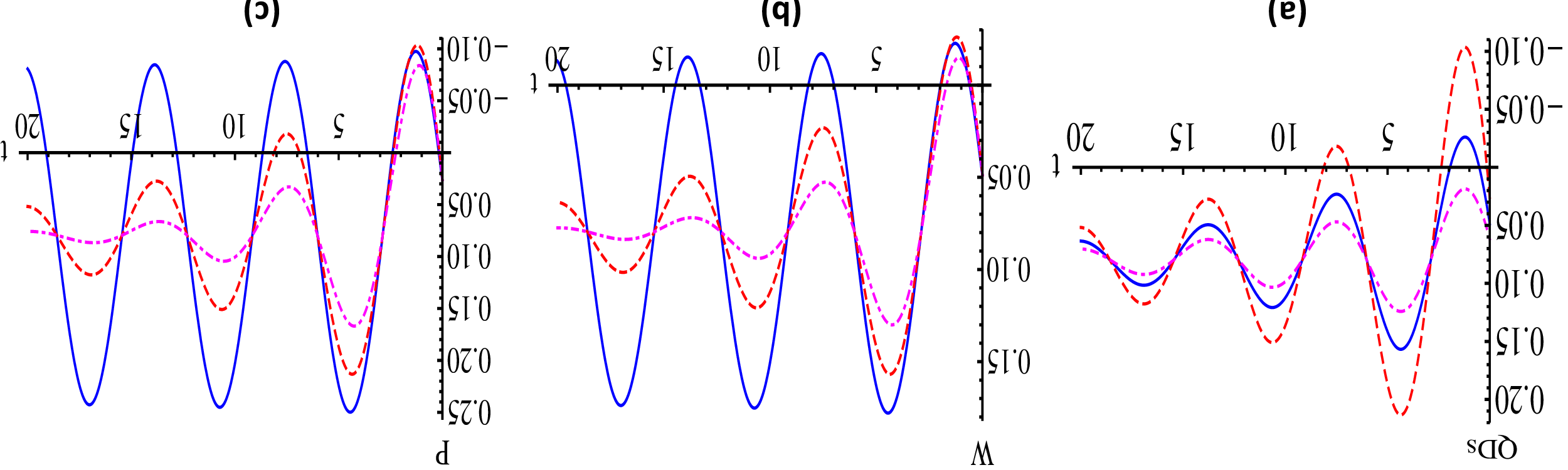} 
\par\end{centering}

\caption{\label{fig:QND-single}(Color online) Plot (a) shows the variation
of all QDs with time ($t$) for single spin-$\frac{1}{2}$ atomic
coherent state in the presence of QND noise with bath parameters $\gamma_{0}=0.1,\,\omega_{c}=100,$
squeezing parameters $r=0,\, a=0,$ and $\omega=1.0$ at temperature
$T=1$, and $\alpha=\frac{\pi}{2},\,\beta=\frac{\pi}{3},\,\theta=\frac{\pi}{3},\,\phi=\frac{\pi}{4},$
in the units of $\hbar=k_{B}=1$. The smooth (blue) line, dashed (red)
line and dot-dashed (magenta) line correspond to the $W$, $P$ and
$Q$ functions, respectively; (b) and (c) show the variation of $W$
and $P$ functions with time for different temperatures $T=0,\,1$
and $2$ by smooth (blue) lines, dashed (red) lines and dot-dashed
(magenta) lines, respectively.}
\end{figure}

Here, we consider the case of an Ohmic bath for which analytic expressions
for $\gamma\left(t\right)$, both for zero and high temperatures,
can be obtained \citep{QND}. These are functions of the bath parameters
$\gamma_{0}$ and $\omega_{c}$ as well as squeezing parameters $r$
and $\phi$, with $\phi=a\omega$ and $a$ is a constant dependent
on the squeezed bath. Now, using multipole operators, mentioned above,
analytic expressions of the different QDs can be obtained. For example,
we have obtained the $W$ function for a qubit, starting from an atomic
coherent state, in the presence of QND noise as 
\begin{equation}
\begin{array}{lcl}
W\left(\theta,\phi\right) & = & \frac{1}{4\pi}\left(1-\sqrt{3}\cos\alpha\cos\theta+\sqrt{3}e^{-\left(\hbar\omega\right)^{2}\gamma\left(t\right)}\cos\left(\beta+\omega t+\phi\right)\sin\alpha\sin\theta\right),\end{array}\label{eq:wigner-QND}
\end{equation}
while, the corresponding $P$ and $Q$ QDs are obtained as 
\begin{equation}
\begin{array}{lcl}
P\left(\theta,\phi\right) & = & \frac{1}{4\pi}\left(1+3\cos\alpha\cos\theta+3e^{-\left(\hbar\omega\right)^{2}\gamma\left(t\right)}\cos\left(\beta+\omega t+\phi\right)\sin\alpha\sin\theta\right),\end{array}\label{eq:P-QND}
\end{equation}
and 
\begin{equation}
\begin{array}{lcl}
Q\left(\theta,\phi\right) & = & \frac{1}{4\pi}\left(1+\cos\alpha\cos\theta+e^{-\left(\hbar\omega\right)^{2}\gamma\left(t\right)}\cos\left(\beta+\omega t+\phi\right)\sin\alpha\sin\theta\right),\end{array}\label{eq:Q-QND}
\end{equation}
respectively. All the QDs calculated in Eqs. (\ref{eq:wigner-QND})-(\ref{eq:Q-QND})
can be used to get the corresponding noiseless QDs for the same system
and this also serves as a nice consistency check of the calculations.
The variation of the QDs, Eqs. (\ref{eq:wigner-QND})-(\ref{eq:Q-QND}),
for some specific parameter values are shown in Fig. \ref{fig:QND-single}
a-c, where the effect of the presence of noise on the QDs can be easily
observed. Both the $P$ and $W$ functions are found to exhibit negative
values indicative of quantumness in the system. Also, in Fig. \ref{fig:QND-single}
b-c, we can see that with an increase in temperature $T$, the QDs
tend to become less negative, which is an indicator of a move towards
classicality, as expected. Interestingly, in Fig. \ref{fig:QND-single}
a, we do not observe any zero of the $Q$ function which implies that
the $Q$ function does not show any signature of nonclassicality in
this particular case. The oscillatory nature of the QDs for atomic coherent state when subjected to 
QND noise can be attributed to the purely dephasing effect of the QND interaction. That is, this process involves
decoherence without any dissipation. At temperature $T=0$ decoherence is minimal and hence an oscillatory pattern is
observed in the depicted time scale. With increase in $T$, resulting in increase in the influence of decoherence,
these oscillations gradually decrease.

\subsection{Atomic coherent state in SGAD noise}

\begin{figure}
\begin{centering}
\includegraphics[angle=-90,scale=0.65]{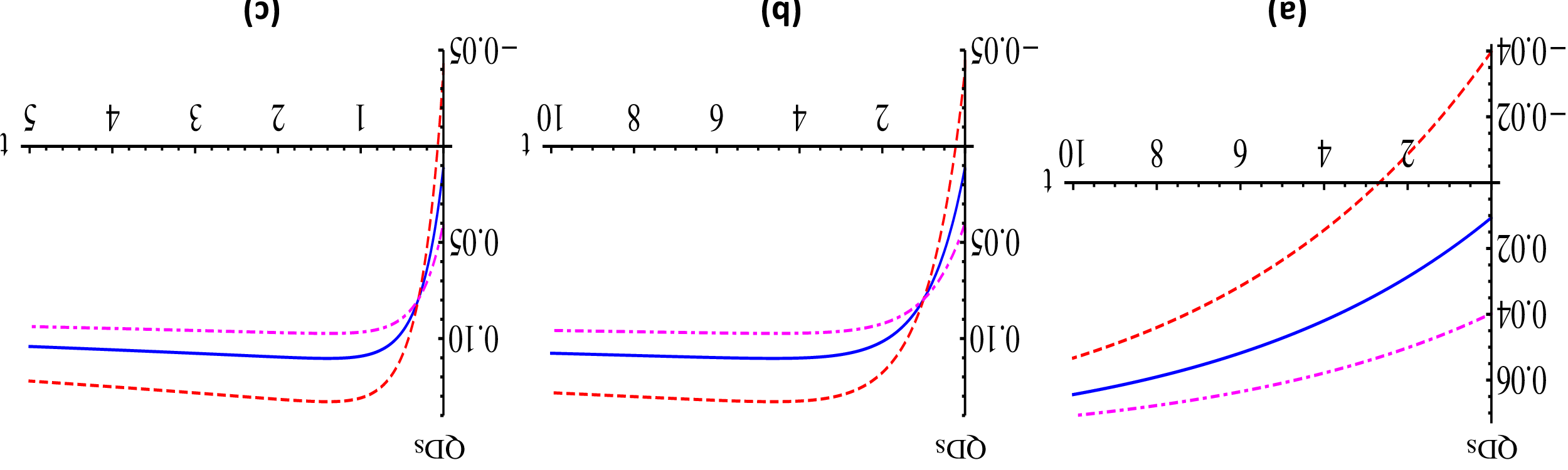} 
\par\end{centering}

\caption{\label{fig:SGAD-ACS}(Color online) The variation of all the distribution
functions with time ($t$) is shown together for a single spin-$\frac{1}{2}$
atomic coherent state in the presence of the SGAD noise for zero bath
squeezing angle in units of $\hbar=k_{B}=1$, with $\omega=1.0,\,\gamma_{0}=0.05,$
and $\alpha=\frac{\pi}{2},\,\beta=\frac{\pi}{3},\,\theta=\frac{\pi}{2},\,\phi=\frac{\pi}{3}.$
In (a) the variation with time is shown for temperature $T=3.0$ in
the absence of squeezing parameter, i.e., $r=0$. In (b) the effect
of the change in squeezing parameter for same temperature, i.e., $T=3.0$
is shown by using the squeezing parameter $r=1.0$, keeping all the
other values as same as that used in (a). Further, in (c) keeping
$r=1.0$ as in (b), the temperature is increased to $T=10$ to show
the effect of variation in T. In (c) time is varied only up to $t=5$
to emphasize the effect of temperature. In all the three plots, smooth
(blue), dashed (red) and dotted-dashed (magenta) lines correspond
to the $W$, $P$ and $Q$ functions, respectively.}
\end{figure}

Now, we take up a spin $j=\frac{1}{2}$, starting from an atomic coherent
state, given by Eq. (\ref{eq:density-matrix}), evolving under a Squeezed
Generalized Amplitude Damping (SGAD) channel, incorporating the effects
of dissipation and bath squeezing and which includes the well known
amplitude damping (AD) and generalized amplitude damping (GAD) channels
as special cases. The Kraus operators of the SGAD channel are \citep{SGAD}
\[
\begin{array}{lcl}
E_{0} & = & \sqrt{p}\left[\begin{array}{cc}
\sqrt{1-\lambda\left(t\right)} & 0\\
0 & 1
\end{array}\right],\end{array}
\]
\[
\begin{array}{lcl}
E_{1} & = & \sqrt{p}\left[\begin{array}{cc}
0 & 0\\
\sqrt{\lambda\left(t\right)} & 0
\end{array}\right],\end{array}
\]
\[
\begin{array}{lcl}
E_{2} & = & \sqrt{1-p}\left[\begin{array}{cc}
\sqrt{1-\mu\left(t\right)} & 0\\
0 & \sqrt{1-\nu\left(t\right)}
\end{array}\right],\end{array}
\]
\begin{equation}
\begin{array}{lcl}
E_{3} & = & \sqrt{1-p}\left[\begin{array}{cc}
0 & \sqrt{\nu\left(t\right)}\\
\sqrt{\mu\left(t\right)}e^{-i\xi\left(t\right)} & 0
\end{array}\right],\end{array}\label{eq:SGAD-kraussoperators2-1}
\end{equation}
where $\lambda=\frac{1}{p}\left\{ 1-\left(1-p\right)\left[\mu+\nu\right]-\exp\left(-\gamma_{0}\left(2N+1\right)t\right)\right\} ,$
$\mu=\frac{2N+1}{2N\left(1-p\right)}\frac{\sinh^{2}\left(\gamma_{0}at/2\right)}{\sinh^{2}\left(\gamma_{0}\left(2N+1\right)t/2\right)}\exp\left(-\frac{\gamma_{0}}{2}\left(2N+1\right)t\right),$
and $\nu=\frac{N}{\left(1-p\right)\left(2N+1\right)}\left\{ 1-\exp\left(-\gamma_{0}\left(2N+1\right)t\right)\right\} .$
Here, for convenience we have omitted the time dependence in the argument
of different time dependent parameters (e.g., $\lambda(t)$, $\mu(t)$,
$\nu(t)$, etc.) in the Kraus operators of SGAD noise. Here, $\gamma_{0}$
is the spontaneous emission rate, $a=\sinh\left(2r\right)\left(2N_{th}+1\right),$
and $N=N_{th}\left\{ \cosh^{2}\left(r\right)+\sinh^{2}\left(r\right)\right\} +\sinh^{2}\left(r\right),$
with $N_{th}=1/\left\{ \exp\left(\hbar\omega/k_{B}T\right)-1\right\} $
being the Planck distribution. Here, $r$ and the bath squeezing angle
($\xi\left(t\right)$) are the bath squeezing parameters. The expression
for $p$ in the above equations has an analytic, though complicated,
expression, and we refer the reader to \citep{SGAD} for details.
Application of the above Kraus operators to the initial state results
in 
\[
\begin{array}{lcl}
\rho\left(t\right) & = & \overset{3}{\underset{i=0}{\Sigma}}E_{i}\left(t\right)\rho\left(0\right)E_{i}^{\dagger}\left(t\right).\end{array}
\]
The state at time $t$ can be obtained as 
\begin{equation}
\rho\left(t\right)=\left[\begin{array}{cc}
\rho_{11} & \rho_{21}^{*}\\
\rho_{21} & 1-\rho_{11}
\end{array}\right],\label{eq:density-matrix-SGAD}
\end{equation}
where 
\[
\begin{array}{lcl}
\rho_{11} & = & \frac{1}{2}\left\{ 1-\mu+\nu-p\left(\lambda-\mu+\nu\right)+\left(-1+\mu+\nu+p\left(\lambda-\mu-\nu\right)\right)\cos\alpha\right\} ,\end{array}
\]
and 
\[
\begin{array}{lcl}
\rho_{21} & = & \frac{1}{2}\sin\alpha\left\{ \left(1-p\right)\sqrt{\mu\nu}e^{-i\left(\beta+\xi\right)}+p\sqrt{1-\lambda}e^{i\beta}+\left(1-p\right)\sqrt{\left(1-\mu\right)\left(1-\nu\right)}e^{i\beta}\right\} .\end{array}
\]
Using this density matrix, we can calculate the evolution of the different
QDs, in a manner similar to the previous example of evolution under
QND channel, leading to 
\begin{equation}
\begin{array}{lcl}
W\left(\theta,\phi\right) & = & \frac{1}{4\pi}\left[1+\sqrt{3}\left\{ -\mu+\nu-p\left(\lambda-\mu+\nu\right)+\left(-1+\mu+\nu+p\left(\lambda-\mu-\nu\right)\right)\cos\alpha\right\} \cos\theta\right.\\
 & + & \sqrt{3}\left(\left\{ p\sqrt{1-\lambda}+\left(1-p\right)\sqrt{\left(1-\mu\right)\left(1-\nu\right)}\right\} \cos\left(\beta+\phi\right)\right.\\
 & + & \left.\left.\left(1-p\right)\sqrt{\mu\nu}\cos\left(\beta+\xi-\phi\right)\right)\sin\alpha\sin\theta\right];
\end{array}\label{eq:Wigner-SGAD}
\end{equation}
\begin{equation}
\begin{array}{lcl}
P\left(\theta,\phi\right) & = & \frac{1}{4\pi}\left[1-3\left\{ -\mu+\nu-p\left(\lambda-\mu+\nu\right)+\left(-1+\mu+\nu+p\left(\lambda-\mu-\nu\right)\right)\cos\alpha\right\} \cos\theta\right.\\
 & + & 3\left(\left\{ p\sqrt{1-\lambda}+\left(1-p\right)\sqrt{\left(1-\mu\right)\left(1-\nu\right)}\right\} \cos\left(\beta+\phi\right)\right.\\
 & + & \left.\left.\left(1-p\right)\sqrt{\mu\nu}\cos\left(\beta+\xi-\phi\right)\right)\sin\alpha\sin\theta\right];
\end{array}\label{eq:P-SGAD}
\end{equation}
and 
\begin{equation}
\begin{array}{lcl}
Q\left(\theta,\phi\right) & = & \frac{1}{4\pi}\left[1-\left\{ -\mu+\nu-p\left(\lambda-\mu+\nu\right)+\left(-1+\mu+\nu+p\left(\lambda-\mu-\nu\right)\right)\cos\alpha\right\} \cos\theta\right.\\
 & + & \left(\left\{ p\sqrt{1-\lambda}+\left(1-p\right)\sqrt{\left(1-\mu\right)\left(1-\nu\right)}\right\} \cos\left(\beta+\phi\right)\right.\\
 & + & \left.\left.\left(1-p\right)\sqrt{\mu\nu}\cos\left(\beta+\xi-\phi\right)\right)\sin\alpha\sin\theta\right].
\end{array}\label{eq:Q-SGAD}
\end{equation}
The variation of all the QDs with time ($t$) for some specific values
of the parameters is depicted in Fig. \ref{fig:SGAD-ACS}, which incorporates
both temperature and squeezing. A comparison of the Figs. \ref{fig:SGAD-ACS}
a and \ref{fig:SGAD-ACS} b brings out the effect of squeezing on
the evolution of QDs. Further, it is easily observed that with the
increase in $T$, the quantumness reduces. An important point to notice
here, is that if we make the noise parameters zero, i.e., in the absence
of noise, the different QDs given by Eqs. (\ref{eq:Wigner-SGAD})-(\ref{eq:Q-SGAD}),
reduce to a form exactly equal to the corresponding noiseless QDs
obtained for QND evolutions (Eqs. (\ref{eq:wigner-QND})-(\ref{eq:Q-QND})).
Also, results for generalized amplitude damping channel can be obtained
in the limit of vanishing squeezing, i.e., for $\mu\left(t\right)=0$
and $\lambda\left(t\right)=\nu\left(t\right)$, while corresponding
results for QDs under evolution of an amplitude damping channel can
be obtained by further setting $T=0$, and $p=1$. Further, it would be apt to mention here that the 
oscillatory nature of the QDs for an atomic coherent state evolving under QND noise is not seen here. 
This is consistent with the fact that the SGAD noise is dissipative in nature, involving decoherence along with
dissipation.

\section{{\normalsize{{QDs for multiqubit systems undergoing QND and dissipative
evolutions}}}}

Now, we wish to study the evolution of QDs for some interesting two
and three qubit systems under general open system evolutions. We will
also take up the well known $N$-qubit Dicke model. In each case,
we study the nonclassicality exhibited by the system under consideration.

\subsection{Two qubits in the presence of QND noise}

\begin{figure}
\centering{}\includegraphics[angle=-90,scale=0.5]{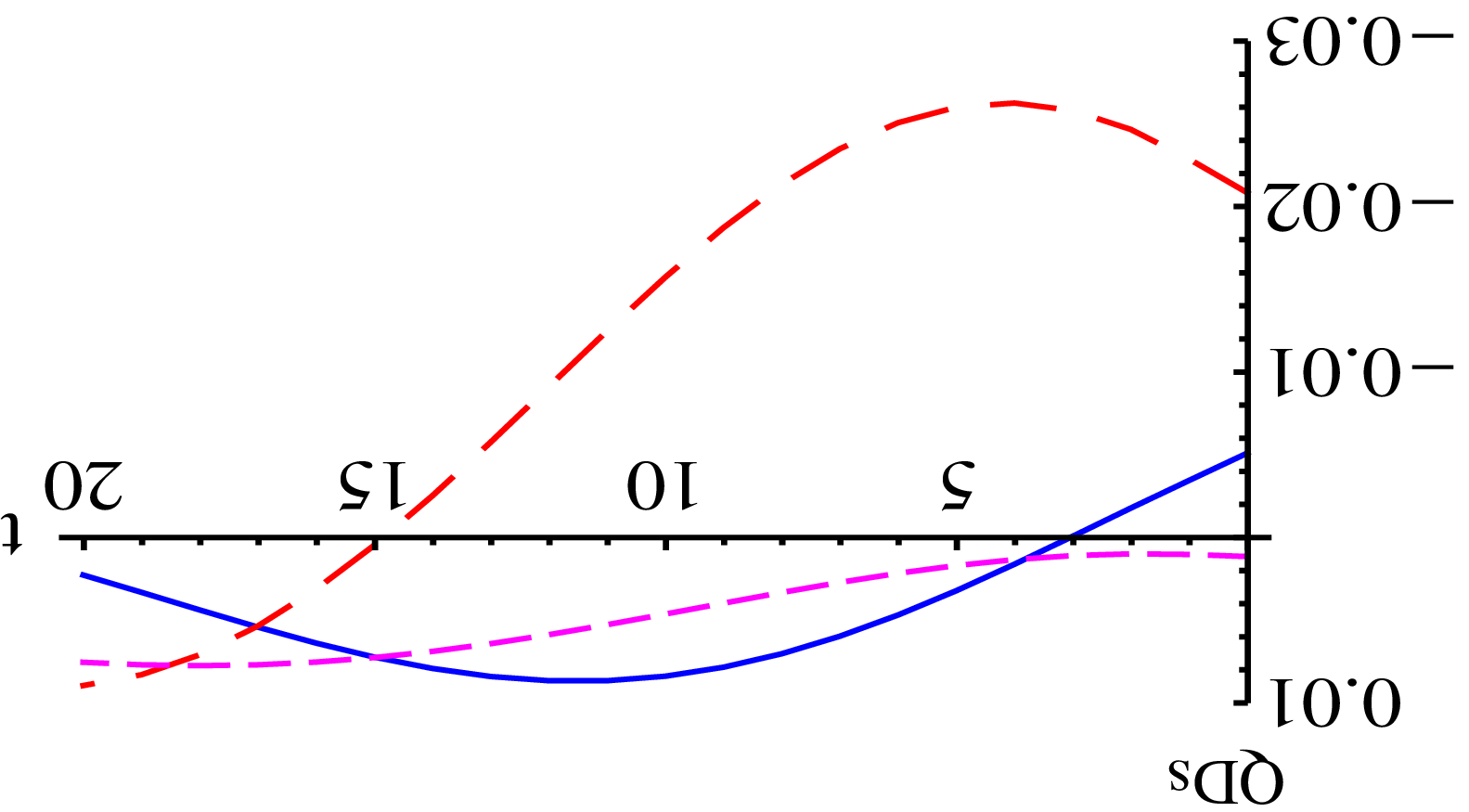}\caption{\label{fig:QND-2qubit}(Color online) The plot shows the variation
of all the QDs with time for two qubits undergoing QND evolution with
$\gamma_{0}=0.01,\,\omega_{c}=100,$ $k.r_{ab}=0.05,$ squeezing parameters
$r=0.05,\, a=0,$ which implies zero squeezing angle, and temperature
$T=2$ with $\theta_{1}=\frac{\pi}{3},\,\theta_{2}=\frac{\pi}{4},\,\phi_{1}=\pi,\,\phi_{2}=\frac{\pi}{3}$
in units of $\hbar=k_{B}=1$. The smooth (blue), large dashed (red),
and small dashed (magenta) lines correspond to $W$, $P$ and $Q$
functions, respectively.}
\end{figure}

The density matrix for a system of two qubits in QND interaction with
a squeezed thermal bath, as obtained in Ref. \citep{2qubit-QND},
is 
\begin{equation}
\begin{array}{lcl}
\rho_{\left\{ i_{n},j_{n}\right\} }^{s}\left(t\right) & = & \exp\left[i\left\{ \Theta_{\left\{ i_{n},j_{n}\right\} }\left(t\right)-\Lambda_{\left\{ i_{n},j_{n}\right\} }\left(t\right)\right\} \right]\exp\left[-\Gamma_{\left\{ i_{n},j_{n}\right\} }^{sq}\left(t\right)\right]\rho^{s}\left(0\right),\end{array}\label{eq:densMat-Lqub-QND}
\end{equation}
where $\rho_{\left\{ i_{n},j_{n}\right\} }^{s}\left(t\right)$ is
the two-qubit reduced density matrix obtained by tracing out the bath
(reservoir) degrees of freedom and has the matrix representation $\left\langle i_{2},i_{1}\right|\rho^{s}\left(t\right)\left|j_{2},j_{1}\right\rangle $,
and $\left\{ i_{n},j_{n}\right\} $ stands for $i_{1},j_{1};i_{2},j_{2}$.
In this model, the system-bath coupling is dependent upon the position
of the qubit, resulting in the classification of the dynamics into
two regimes: (a) Localized model, where the inter-qubit spacing is
greater than or of the order of the length scale set by the bath,
and (b) Collective model, where the qubits are close enough to experience
the same bath \citep{2qubit-QND}. Here, for the sake of brevity,
we will provide details of the localized model only. The terms $\Theta_{\left\{ i_{n},j_{n}\right\} }\left(t\right)$,
$\Lambda_{\left\{ i_{n},j_{n}\right\} }\left(t\right)$ and $\Gamma_{\left\{ i_{n},j_{n}\right\} }^{sq}\left(t\right)$
have different expressions in the localized and collective models.
The superscript $sq$ indicates that the bath starts in a squeezed
thermal initial state. For convenience, the two particle index in
Eq. (\ref{eq:densMat-Lqub-QND}) is denoted by a single 4-level index
in the following manner: 
\[
-\frac{1}{2},-\frac{1}{2}\equiv0;\;-\frac{1}{2},\frac{1}{2}\equiv1;\;\frac{1}{2},-\frac{1}{2}\equiv2;\;\frac{1}{2},\frac{1}{2}\equiv3.
\]
All the sixteen terms, of the density matrix, can be analytically
calculated for a given initial state $\rho^{s}\left(0\right)$. In
the localized model, considered here, the density matrix is obtained
using the symmetry of the density matrix $\rho^{s}\left(t\right)$,
i.e., symmetries between the matrix elements and hermiticity of the
density matrix, and the expressions of different terms in Eq. (\ref{eq:densMat-Lqub-QND})
\citep{2qubit-QND}. The elements 
\[
\rho_{32}^{s}\left(t\right)=\rho_{23}^{*s}\left(t\right)=\rho_{01}^{s}\left(t\right)=\rho_{10}^{*s}\left(t\right),
\]
are obtained using 
\[
\begin{array}{lcl}
\Theta_{32}\left(t\right) & = & \Theta_{01}\left(t\right)=-\Theta_{23}\left(t\right)=-\Theta_{10}\left(t\right)\\
 & = & \intop_{0}^{\infty}d\omega I\left(\omega\right)S\left(\omega,t\right)\cos\omega t_{s},
\end{array}
\]
and 
\[
\begin{array}{lcl}
\Lambda_{32}\left(t\right) & = & \Lambda_{01}\left(t\right)=-\Lambda_{23}\left(t\right)=-\Lambda_{10}\left(t\right)\\
 & = & -\intop_{0}^{\infty}d\omega I\left(\omega\right)C\left(\omega,t\right)\sin\omega t_{s},
\end{array}
\]
where $I\left(\omega\right)$ is the bath spectral density. In the
Ohmic case considered here, $I\left(\omega\right)=\frac{\gamma_{0}}{\pi}\omega e^{-\omega/\omega_{c}},$
where $\gamma_{0}$ and $\omega_{c}$ are bath parameters. We have
$S\left(\omega,t\right)=\frac{\omega t-\sin\omega t}{\omega^{2}},$
$C\left(\omega,t\right)=\frac{1-\cos\omega t}{\omega^{2}},$ $\omega t_{s}\equiv k.r_{mn},$
where $r_{mn}$ is the inter-qubit spacing and $t_{s}$ is the transit
time introduced for the purpose of expressing the coupling of the
system to its bath in the frequency domain. The diagonal elements
of the density matrix are 
\[
\rho_{aa}^{s}\left(t\right)=\rho_{aa}^{s}\left(0\right),{\rm \; where}\, a=0,1,2,3,
\]
implying an unchanging population, a characteristic of QND evolution.
Also, 
\[
\rho_{21}^{s}\left(t\right)=\rho_{12}^{*s}\left(t\right)=\rho_{12}^{s}\left(t\right),
\]
\[
\rho_{30}^{s}\left(t\right)=\rho_{03}^{*s}\left(t\right)=\rho_{03}^{s}\left(t\right),
\]
i.e., these elements are purely real and for these $\Theta\left(t\right)=0=\Lambda\left(t\right).$
The remaining elements are 
\[
\rho_{31}^{s}\left(t\right)=\rho_{13}^{*s}\left(t\right)=\rho_{02}^{s}\left(t\right)=\rho_{20}^{*s}\left(t\right),
\]
and for their calculation we need 
\[
\begin{array}{lcl}
\Theta_{31}\left(t\right) & = & \Theta_{02}\left(t\right)=-\Theta_{13}\left(t\right)=-\Theta_{20}\left(t\right)\\
 & = & \intop_{0}^{\infty}d\omega I\left(\omega\right)S\left(\omega,t\right)\cos\omega t_{s},
\end{array}
\]
and 
\[
\begin{array}{lcl}
\Lambda_{31}\left(t\right) & = & \Lambda_{02}\left(t\right)=-\Lambda_{13}\left(t\right)=-\Lambda_{20}\left(t\right)\\
 & = & \intop_{0}^{\infty}d\omega I\left(\omega\right)C\left(\omega,t\right)\sin\omega t_{s}.
\end{array}
\]
For the determination of all these elements of the density matrix
at time $t$, we also need $\Gamma^{sq}\left(t\right)$ which have
complex expressions and can be seen in \citep{2qubit-QND}. Once the
density matrix at time $t$ is obtained, the corresponding QDs can
be obtained from the prescription discussed above. However, getting
analytic expressions for different QDs, here, is a cumbersome task;
hence we will resort to numerically plotting them. Without loss of
generality, we consider here our initial state to be such that all
the sixteen elements of the density matrix at time $t=0$ are 0.25.
The effect of QND interaction on different QDs obtained for this particular
choice of initial state is shown in Fig. \ref{fig:QND-2qubit}. As
can be seen from the figure, both the $P$ and $W$ functions exhibit
negative values, indicative of quantumness in the system, for some
time after initiation of the evolution, before becoming positive,
due to dephasing caused by the bath. As expected, the $P$ function
is a stronger indicator of quantumness than the $W$ function, while
the $Q$ function is always positive, by construction.

\subsection{Two qubits under dissipative evolution}

Here, we study the evolution of QDs, for two qubit systems, undergoing
dissipative evolution, first interacting with a vacuum bath, $T=0$
and zero bath squeezing, and then under the influence of a squeezed
thermal bath, finite $T$ and bath squeezing. Here, we will make use
of the results worked out in Ref. \citep{squ-ther-bath}.

\subsubsection{Vacuum bath \label{sub:Vacuum-bath}}

The density matrix, in the dressed state basis, can be used for calculating
different QDs. We consider the initial state with one qubit in the
excited state $\left|e_{1}\right\rangle $ and the other in the ground
state $\left|g_{2}\right\rangle $, i.e., $\left|e_{1}\right\rangle \left|g_{2}\right\rangle $.
The two-qubit reduced density matrix is given by 
\begin{equation}
\begin{array}{lcl}
\rho\left(t\right) & = & \left[\begin{array}{cccc}
\rho_{ee}\left(t\right) & \rho_{es}\left(t\right) & \rho_{ea}\left(t\right) & \rho_{eg}\left(t\right)\\
\rho_{es}^{*}\left(t\right) & \rho_{ss}\left(t\right) & \rho_{sa}\left(t\right) & \rho_{sg}\left(t\right)\\
\rho_{ea}^{*}\left(t\right) & \rho_{sa}^{*}\left(t\right) & \rho_{aa}\left(t\right) & \rho_{ag}\left(t\right)\\
\rho_{eg}^{*}\left(t\right) & \rho_{sg}^{*}\left(t\right) & \rho_{ag}^{*}\left(t\right) & \rho_{gg}\left(t\right)
\end{array}\right],\end{array}\label{eq:densitymatrix-vaccumbath}
\end{equation}
where analytic expressions of all the elements of the density matrix
in Eq. (\ref{eq:densitymatrix-vaccumbath}) can be seen from Eqs.
(23)-(32) of Ref. \citep{squ-ther-bath}.

Here, we consider identical qubits. The dynamics involve collective
coherent effects due to the multiqubit interaction, as well collective
incoherent effects due to dissipative multiqubit interaction with
the bath, and spontaneous emission. Analytic expressions of the corresponding
QDs are very cumbersome, hence we resort to numerically studying the
QDs for some parameters. Values of different parameters are as follows
wavevector and mean frequency $k_{0}=\omega_{0}=1$, spontaneous emission
rate $\Gamma_{j}=0.05$, and $\hat{\mu}\cdot\hat{r}_{ij}=0$, where
$\hat{\mu}$ is equal to the unit vector along the atomic transition
dipole moment and $\hat{r}_{ij}$ is the interatomic distance. Considering
the initial state with $\rho_{ee}\left(0\right)=\rho_{gg}\left(0\right)=\rho_{es}\left(0\right)=\rho_{ea}\left(0\right)=\rho_{eg}\left(0\right)=\rho_{sg}\left(0\right)=\rho_{ag}\left(0\right)=0$,
and $\rho_{ss}\left(0\right)=\rho_{aa}\left(0\right)=\rho_{sa}\left(0\right)=0.5$,
the $W$, $P$, and $Q$ functions are calculated.

\begin{figure}
\centering{}\includegraphics[angle=-90,scale=0.8]{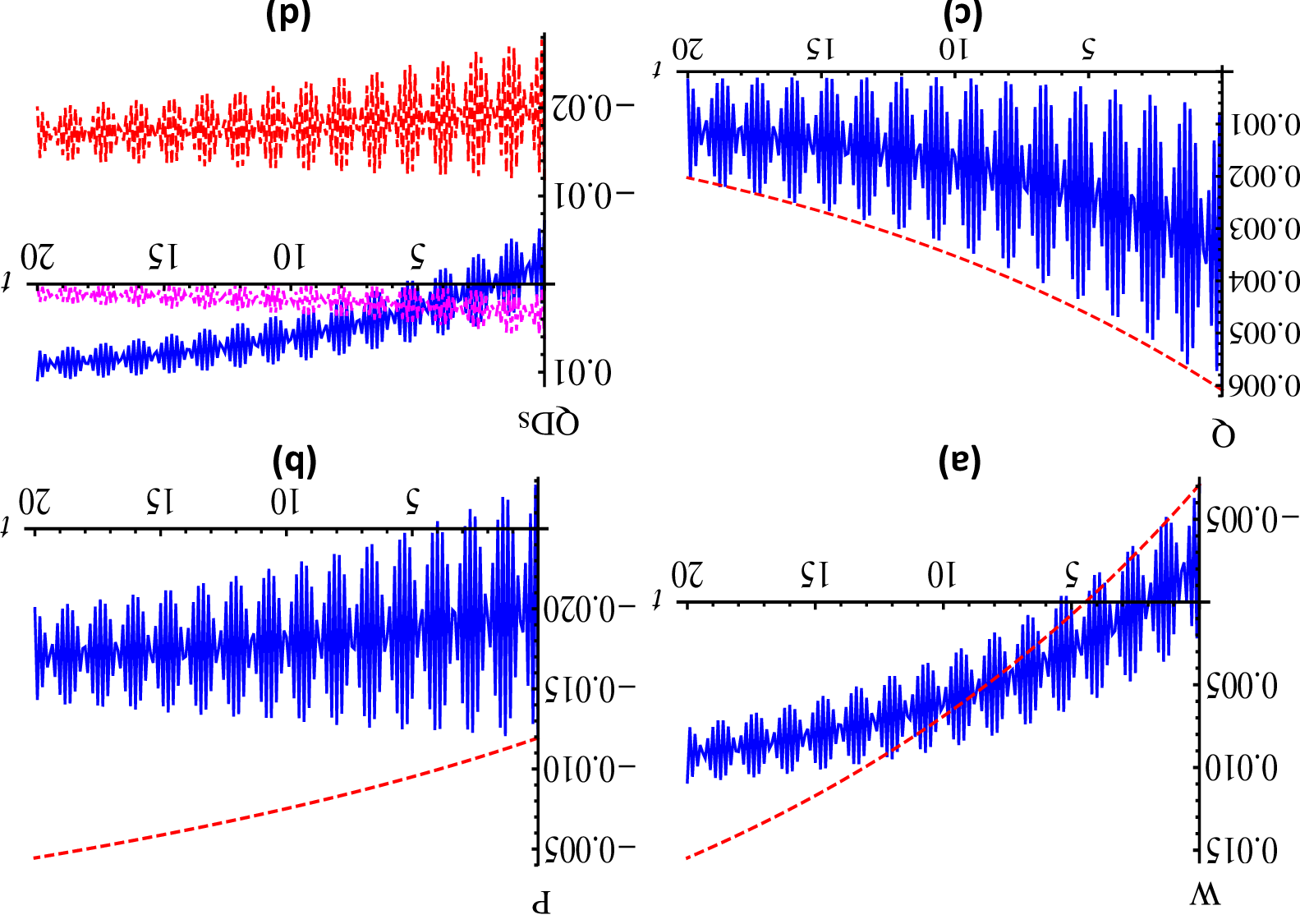}\caption{\label{fig:Vacuum-bath}(Color online) The variation of the $W$,
$P$, and $Q$ functions with time is shown in (a)-(c) for the two-qubit
state, in the presence of vacuum bath, with $\theta_{1}=\frac{\pi}{8},\,\theta_{2}=\frac{\pi}{3},\,\phi_{1}=\frac{\pi}{4},\,\phi_{2}=\frac{\pi}{4}$
with the inter-qubit spacing $r_{12}=0.05$ (smooth blue line), and
$r_{12}=2.0$ (red dashed line). In (d) all the QDs, varying with
time, are plotted together with inter-qubit spacing $r_{12}=0.05$.
Here the $P$ function is seen to be negative for all times shown,
while the $W$ function is negative only for sometime, while the $Q$
function is always positive.}
\end{figure}

\begin{figure}
\centering{}\includegraphics[angle=-90,scale=0.8]{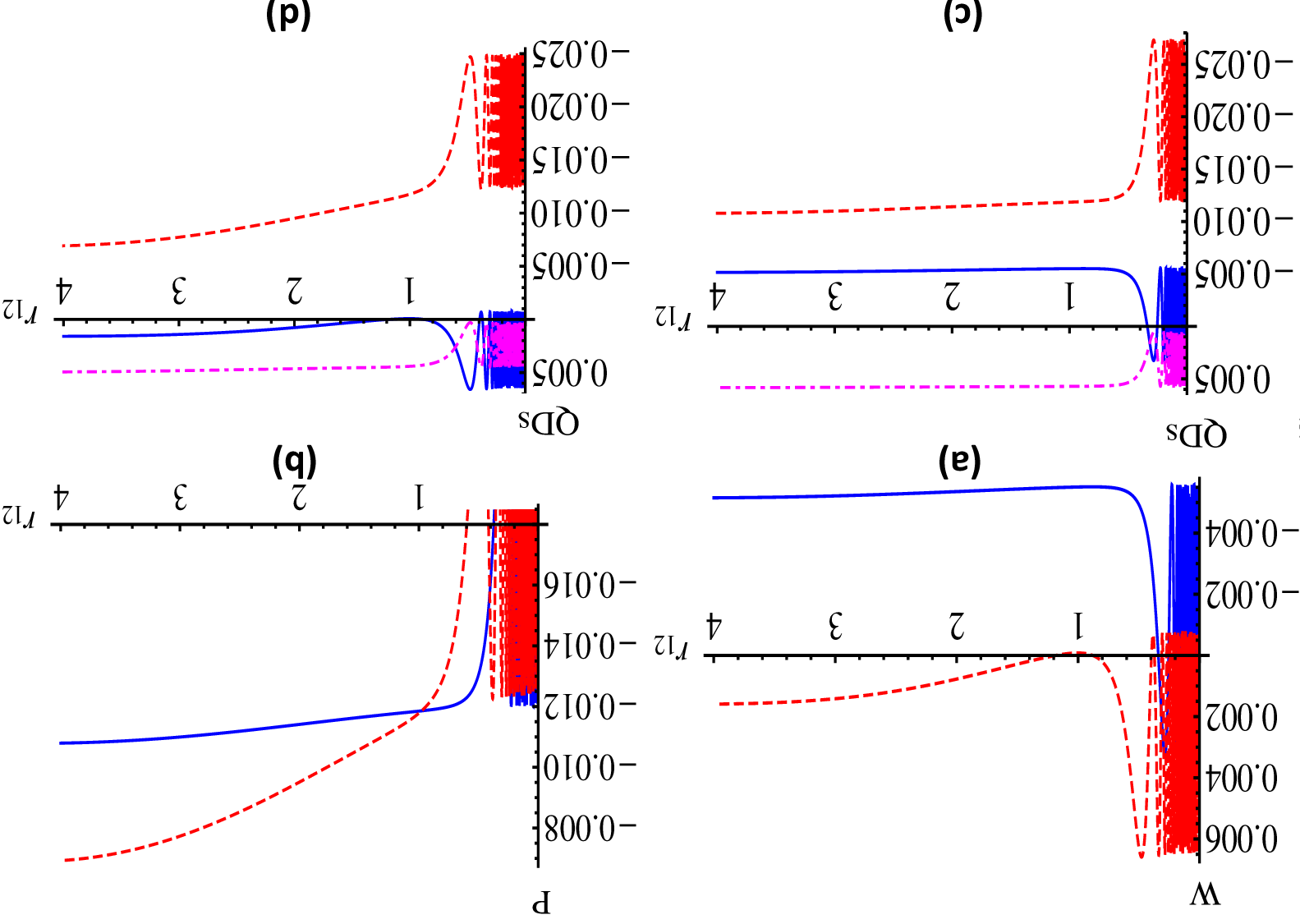}\caption{\label{fig:Vacuumbath-2}(Color online) (a) and (b) depict the $W$
and $P$ functions for the two qubit state, interacting with a vacuum
bath, as a function of the inter-qubit spacing at $t=1$ (smooth blue
line), and $t=5$ (red dashed line). In (c) and (d), $W$ (smooth
blue line), $P$ (red dashed line) and $Q$ (magenta dot-dashed line)
QDs are plotted together, depicting their variation with inter-qubit
spacing at time $t=1$ and $t=5$, respectively. For all the plots
$\theta_{1}=\frac{\pi}{8},\,\theta_{2}=\frac{\pi}{3},\,\phi_{1}=\frac{\pi}{4},\,{\rm and}\,\phi_{2}=\frac{\pi}{4}$.}
\end{figure}

The variation of the different QDs is depicted in Figs. \ref{fig:Vacuum-bath}
and \ref{fig:Vacuumbath-2}. Figs. \ref{fig:Vacuum-bath} a and \ref{fig:Vacuum-bath}
b show the negative values of $W$ function and $P$ function for
some and for all times, respectively. The $Q$ function exhibits a
decaying pattern. These features are reinforced in the last plot of
the figure, where all the QDs are plotted together. In Fig. \ref{fig:Vacuumbath-2},
various QDs are plotted with respect to the inter-qubit distance.
In the collective regime, $r_{12}\ll1$, the QDs exhibit an oscillatory
behavior, in consonance with the general behavior in this regime \citep{squ-ther-bath}.
Also, for the chosen parameters, the $P$ function is always negative,
while the $W$ function is negative for $t=1$, but becomes positive
for a longer time $t=5$, due to the dissipative influence of the
bath.

\subsubsection{Squeezed thermal bath}

\begin{figure}
\begin{centering}
\includegraphics[angle=-90,scale=0.6]{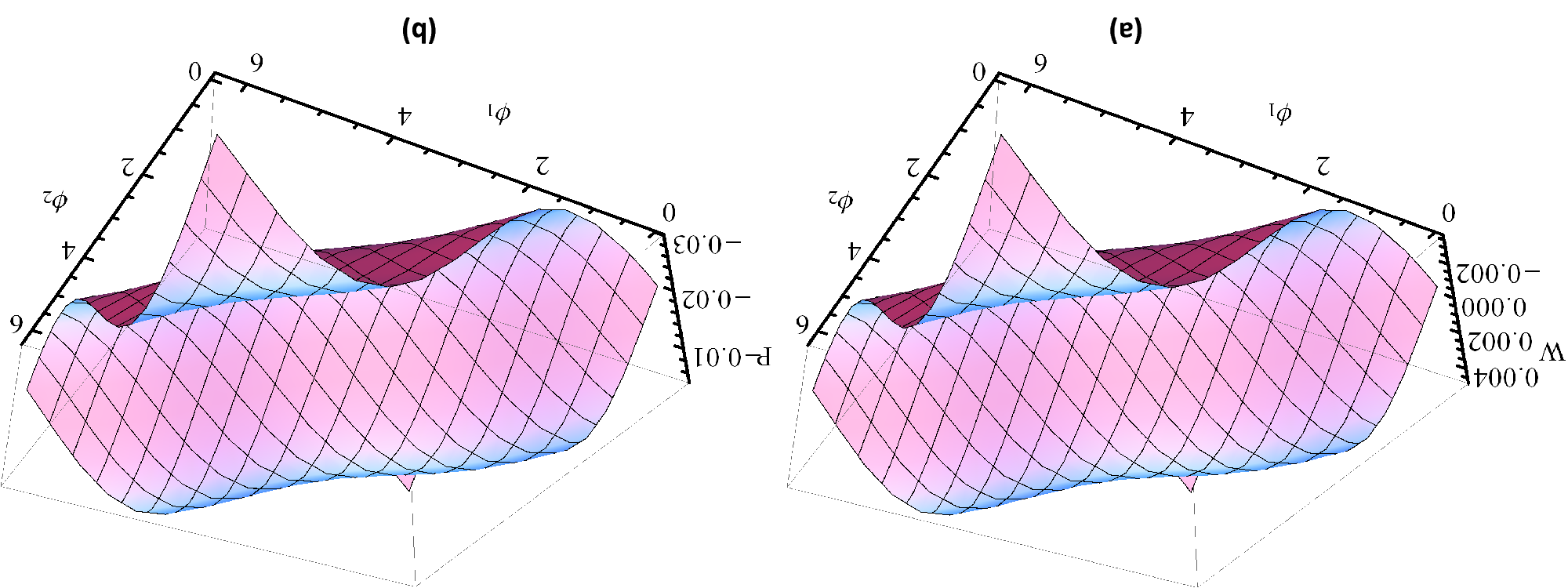} 
\par\end{centering}

\caption{\label{fig:Squeezed-thermal-ph}(Color online) $W$ and $P$ functions
were able to detect the nonclassicality in the presence of squeezed
thermal bath at time $t=2.0$ with $\theta_{1}=\frac{\pi}{4},\,\theta_{2}=\frac{\pi}{8}$,
at temperature $T=1.0,$ squeezing parameter $r=0.5,$ and interatomic
distance $kr_{12}=0.05$; all the remaining parameters are as mentioned
above. The $Q$ function failed to detect nonclassicality. The $P$
function shows nonclassicality for all values of $\phi_{1}$ and $\phi_{2}$
as it is negative for all values of the chosen parameters.}
\end{figure}

\begin{figure}
\centering{}\includegraphics[angle=-90,scale=0.8]{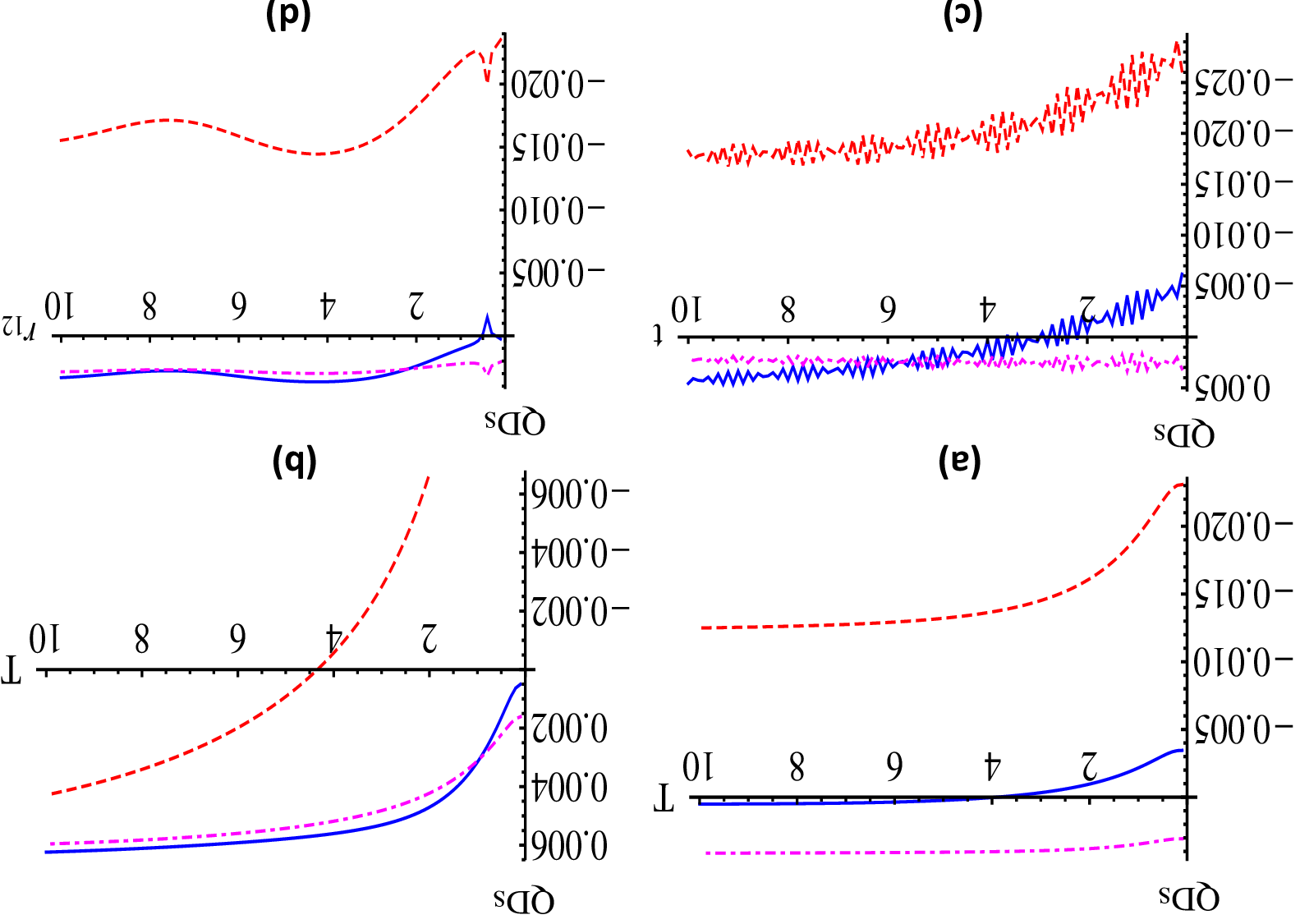} \caption{\label{fig:Squeezed-thermal}(Color online) Different QDs, evolving
under the influence of a squeezed thermal bath, are plotted with $\theta_{1}=\frac{\pi}{4},\,\theta_{2}=\frac{\pi}{8},\,\phi_{1}=\frac{\pi}{6},\,\phi_{2}=\frac{\pi}{8}$,
where smooth (blue), dashed (red) and dot-dashed (magenta) lines correspond
to the $W$, $P$ and $Q$ functions, respectively. (a) and (b) are
for temperature variation at time $t=2.0$ in two cases, (a) $r=0.5,\, kr_{12}=0.08$
(collective model), and (b) $r=-0.5,\, kr_{12}=1.5$ (localized model).
In (c) all the QDs are plotted with respect to time with $T=1.0,\, r=0.1,\, kr_{12}=0.05$.
And (d) shows the variation of all the QDs with $r_{12}$ for $T=1.0,\, t=2.0,\, r=0.5$.}
\end{figure}

Let us consider the evolution of same initial state, as in the presence
of the vacuum bath at time $t=0$, but now evolving under the influence
of a squeezed thermal bath with finite $T$ and $r$. Similar to the
case of the vacuum bath, for certain values of the parameters in the
density matrix, we can calculate different QDs. The elements of the
density matrix in the presence of a squeezed thermal bath can be obtained
as in Eqs. (33)-(40) in Ref. \citep{squ-ther-bath}. For simplicity,
we take here the squeezing angle $\Phi=0$, and all other parameters
are same as in the case of the vacuum bath, i.e., $k_{0}=\omega_{0}=1$,
$\Gamma_{j}=0.05$, and $\hat{\mu}\cdot\hat{r}_{ij}=0$. From the
density matrix of the evolved state, the various QDs can be obtained,
which once more due to their cumbersome nature are studied numerically.
The behavior of the different QDs is shown, for different parameters,
in Figs. \ref{fig:Squeezed-thermal-ph} and \ref{fig:Squeezed-thermal}.

In Figs. \ref{fig:Squeezed-thermal-ph}, three dimensional plots of
$W$ and $P$ QDs are shown with respect to the azimuthal angles.
The $P$ function exhibits negative values for all values of the parameters
chosen, while the $W$ function does so for a restricted set of values.
All these reiterate the quantumness of the state studied. From Fig.
\ref{fig:Squeezed-thermal} a-b, the effect of finite bath squeezing
$r$ and $T$ on the evolution of the QDs can be seen. In particular,
with an increase in $T$, the QDs, both $P$ and $W$, which were
earlier exhibiting negative values start becoming positive, a clear
indicator of a quantum to classical transition. Fig. \ref{fig:Squeezed-thermal}
c and d, showing the behavior of the QDs with respect to time and
inter-qubit separations, are also along the expected lines. The nature of the QDs for 
small interqubit spacing, as seen here, is consistent with that of the same state subjected to dissipative interaction
with a vacuum bath. Due to increase in temperature the oscillations observed in the QDs for small interqubit 
spacing, for the case of interaction with the vacuum bath in Fig. \ref{fig:Vacuumbath-2}d, decreases in the present
scenario of a squeezed thermal bath interaction and depicted in Fig. \ref{fig:Squeezed-thermal}d. Further increase in 
temperature flattens the peak, observed for small inter-qubit spacing, 
in Fig. \ref{fig:Squeezed-thermal}d.

\subsubsection{EPR singlet state in an amplitude damping (AD) channel}

Now, we take an initially entangled two qubit, EPR singlet, state
\citep{EPR}. The evolution of this state is studied assuming independent
action of an amplitude damping (AD) channel on each qubit. Such a
scenario could be envisaged in a quantum memory net with the qubits
being its remote components, subject locally to the AD noise \citep{eberly}.
Using Kraus operators of an AD channel 
\[
\begin{array}{lcl}
E_{0} & = & \left[\begin{array}{cc}
\sqrt{1-\lambda\left(t\right)} & 0\\
0 & 1
\end{array}\right],\end{array}
\]
\begin{equation}
\begin{array}{lcl}
E_{1} & = & \left[\begin{array}{cc}
0 & 0\\
\sqrt{\lambda\left(t\right)} & 0
\end{array}\right],\end{array}\label{eq:AD-kraussoperators}
\end{equation}
where $\begin{array}{lcl}
\lambda\left(t\right) & = & 1-e^{-\gamma_{0}t},\end{array}$ where $\gamma_{0}$ is the spontaneous emission rate, and assuming
that the two qubits, of the singlet, are independent and do not have
any interaction, the Kraus operators for the action of AD channels,
one on each spin, can be modeled as 
\[
\begin{array}{lcl}
K_{1} & = & E_{0}\left(A\right)\otimes E_{0}\left(B\right),\\
K_{2} & = & E_{0}\left(A\right)\otimes E_{1}\left(B\right),\\
K_{3} & = & E_{1}\left(A\right)\otimes E_{0}\left(B\right),\\
K_{4} & = & E_{1}\left(A\right)\otimes E_{1}\left(B\right),
\end{array}
\]
where $A$ and $B$ stand for the first and second qubits (spins)
comprising the singlet, respectively. From the form of Kraus operators
(\ref{eq:AD-kraussoperators}) and assuming $\lambda_{A}=\lambda_{B}=\lambda$,
we have 
\[
\begin{array}{lcl}
K_{1} & = & \left[\begin{array}{cccc}
1-\lambda & 0 & 0 & 0\\
0 & \sqrt{1-\lambda} & 0 & 0\\
0 & 0 & \sqrt{1-\lambda} & 0\\
0 & 0 & 0 & 1
\end{array}\right],\end{array}
\]
\[
\begin{array}{lcl}
K_{2} & = & \left[\begin{array}{cccc}
0 & 0 & 0 & 0\\
\sqrt{\lambda\left(1-\lambda\right)} & 0 & 0 & 0\\
0 & 0 & 0 & 0\\
0 & 0 & \sqrt{\lambda} & 0
\end{array}\right],\end{array}
\]
\[
\begin{array}{lcl}
K_{3} & = & \left[\begin{array}{cccc}
0 & 0 & 0 & 0\\
0 & 0 & 0 & 0\\
\sqrt{\lambda\left(1-\lambda\right)} & 0 & 0 & 0\\
0 & \sqrt{\lambda} & 0 & 0
\end{array}\right],\end{array}
\]
\begin{equation}
\begin{array}{lcl}
K_{4} & = & \left[\begin{array}{cccc}
0 & 0 & 0 & 0\\
0 & 0 & 0 & 0\\
0 & 0 & 0 & 0\\
\lambda & 0 & 0 & 0
\end{array}\right].\end{array}\label{eq:AD-Krauss-2qubit}
\end{equation}
The density matrix of the singlet state at time $t$, under the action
of the above channel is 
\[
\begin{array}{lcl}
\rho\left(t\right) & = & \overset{4}{\underset{i=1}{\Sigma}}K_{i}\left(t\right)\rho\left(0\right)K_{i}^{\dagger}\left(t\right),\end{array}
\]
where $\rho\left(0\right)=\left|\phi\right\rangle \left\langle \phi\right|$,
and $\begin{array}{lcl}
\left|\phi\right\rangle  & = & \frac{1}{\sqrt{2}}\left(\left|\frac{1}{2},-\frac{1}{2}\right\rangle -\left|-\frac{1}{2},\frac{1}{2}\right\rangle \right),\end{array}$ is the initial state at time $t=0$. Hence, at time $t$ the evolved
density matrix is 
\begin{equation}
\begin{array}{lcl}
\rho\left(t\right) & = & \left[\begin{array}{cccc}
0 & 0 & 0 & 0\\
0 & \frac{1}{2}\left(1-\lambda\right) & -\frac{1}{2}\left(1-\lambda\right) & 0\\
0 & -\frac{1}{2}\left(1-\lambda\right) & \frac{1}{2}\left(1-\lambda\right) & 0\\
0 & 0 & 0 & \lambda
\end{array}\right].\end{array}\label{eq:rho-ADchannel}
\end{equation}

\begin{figure}
\centering{}\includegraphics[angle=-90,scale=0.5]{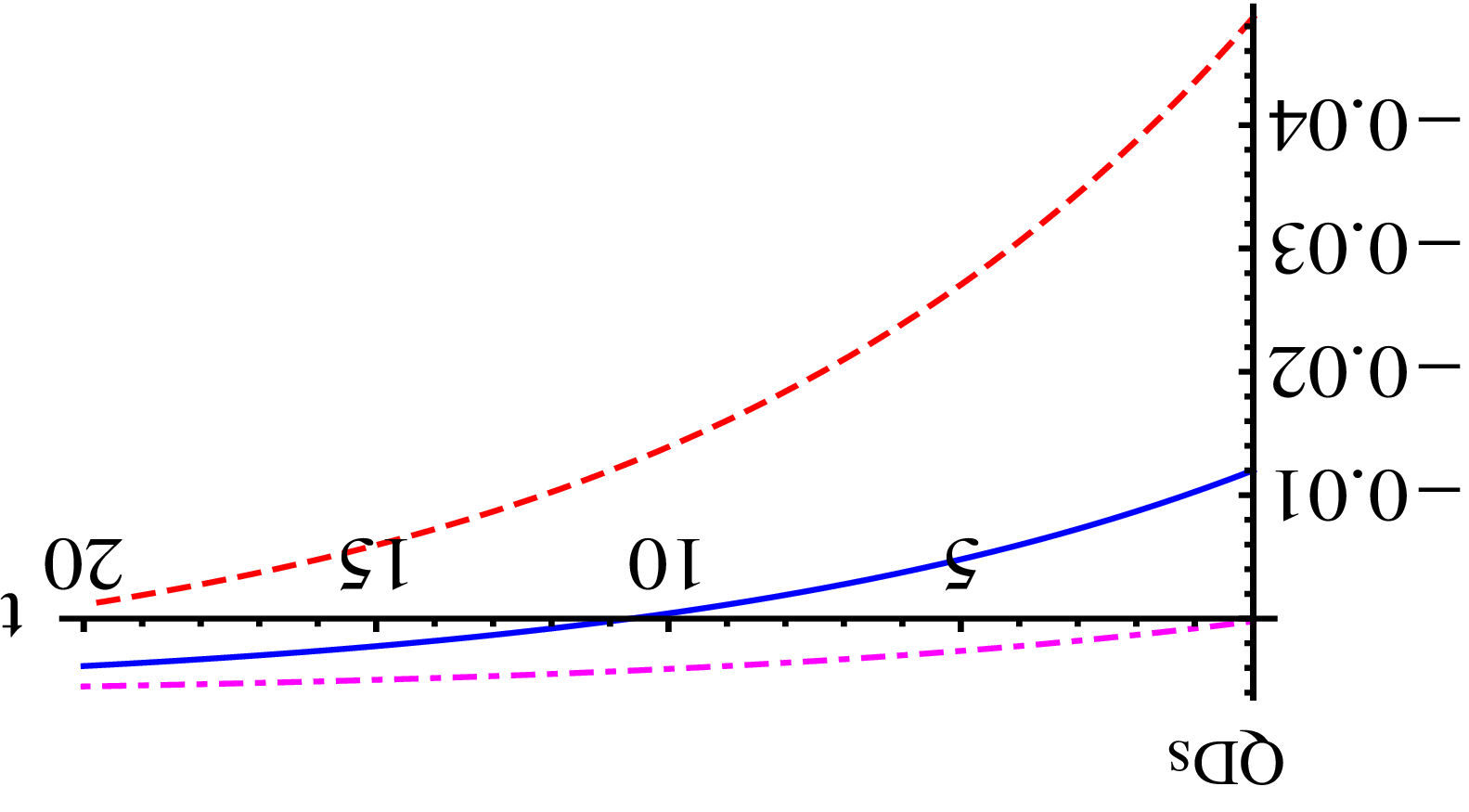}\caption{\label{fig:EPR-AD}(Color online) The variation of all the QDs with
time $t$ for EPR singlet state in the presence of AD channel with
$\gamma_{0}=0.1$ and $\theta_{1}=\frac{\pi}{2},\theta_{2}=\frac{\pi}{2},\phi_{1}=\frac{\pi}{4},\phi_{2}=\frac{\pi}{3}$.
Smooth (blue), dashed (red) and dot-dashed (magenta) lines are for
$W$, $P$ and $Q$ functions, respectively.}
\end{figure}

On the evolved state, represented by the above density matrix, we
may now apply the prescription for obtaining the QDs to yield compact
analytical expressions of the various QDs. Specifically, the $W$
function is obtained as 
\begin{equation}
\begin{array}{lcl}
W\left(\theta_{1},\phi_{1},\theta_{2},\phi_{2}\right) & = & \frac{1}{16\pi^{2}}\left[\lambda\left\{ 1+3\cos\theta_{1}\cos\theta_{2}-\sqrt{3}\left(\cos\theta_{1}+\cos\theta_{2}\right)\right\} \right.\\
 & + & \left.\left(1-\lambda\right)\left\{ 1-3\cos\theta_{1}\cos\theta_{2}-3\sin\theta_{1}\sin\theta_{2}\cos\left(\phi_{1}-\phi_{2}\right)\right\} \right],
\end{array}\label{eq:Wigner-AD-EPR}
\end{equation}
while the $P$ function is 
\begin{equation}
\begin{array}{lcl}
P\left(\theta_{1},\phi_{1},\theta_{2},\phi_{2}\right) & = & \frac{1}{16\pi^{2}}\left[\lambda\left\{ 1+9\cos\theta_{1}\cos\theta_{2}+3\left(\cos\theta_{1}+\cos\theta_{2}\right)\right\} \right.\\
 & + & \left.\left(1-\lambda\right)\left\{ 1-9\cos\theta_{1}\cos\theta_{2}-9\sin\theta_{1}\sin\theta_{2}\cos\left(\phi_{1}-\phi_{2}\right)\right\} \right],
\end{array}\label{eq:P-AD-EPR}
\end{equation}
and the $Q$ function is 
\begin{equation}
\begin{array}{lcl}
Q\left(\theta_{1},\phi_{1},\theta_{2},\phi_{2}\right) & = & \frac{1}{16\pi^{2}}\left[\lambda\left\{ 1+\cos\theta_{1}\cos\theta_{2}+\left(\cos\theta_{1}+\cos\theta_{2}\right)\right\} \right.\\
 & + & \left.\left(1-\lambda\right)\left\{ 1-\cos\theta_{1}\cos\theta_{2}-\sin\theta_{1}\sin\theta_{2}\cos\left(\phi_{1}-\phi_{2}\right)\right\} \right].
\end{array}\label{eq:Q-AD-EPR}
\end{equation}
The QDs, reported here for an EPR pair (singlet state) evolving under
AD channel exactly match with the corresponding noiseless results
\citep{agarwal,rama}, by setting $\lambda=0$ in the above expressions.
The variation of the different QDs with time is shown in Fig. \ref{fig:EPR-AD}.
The $P$ and $W$ functions are found to show negative values for
a long time, indicative of the perfect initial entanglement in the
system, and finally become positive due to exposure to noise.

As all the QDs, in this case, are symmetric pairwise in $\left(\theta_{1}\leftrightarrow\theta_{2}\right)$
and $\left(\phi_{1}\leftrightarrow\phi_{2}\right)$, hence either
or both of these exchanges would leave the expressions unchanged.
For $\theta_{1}=-\theta_{2}=\frac{\pi}{2}$, and $\phi_{1}=\phi_{2}=0$,
we observe the classically perfect anti-correlation of spins. We can
also observe that for certain angles, viz. $\theta_{1}=\theta_{2}=\frac{\pi}{2}$,
and $\phi_{1}-\phi_{2}=\frac{n\pi}{2},$ where $n$ is an odd integer,
all the QDs become equal to $\frac{1}{\left(4\pi\right)^{2}}$, a
result which remains unaffected by the presence or absence of noise.
Hence, for these settings, the evolution of the QDs becomes noise
independent.

\subsection{Three qubit QDs evolution in an AD channel}

Three qubit entangled states can be classified into two classes (GHZ
and W classes) of quantum states, such that a state of W (GHZ) class
cannot be transformed to a state of GHZ (W) class by using LOCC (local
operation and classical communication). Here, we study both GHZ \citep{ghz}
and W \citep{w} classes of states. To simulate the effect of noise,
we consider the scenario wherein the first qubit is affected by the
AD channel. An arbitrary effect of noise on each subsystem could be thought of as more natural. 
The assumption of only one subsystem affected by amplitude damping noise is consistent with the 
effect of noise considered in, for example, various cryptrographic protocols (\cite{our-cryp} and references therein), 
where it is commonly assumed that the qubits which travel through the channel are affected by noise 
while the channel noise does not affect the qubits to be teleported or not travelling through it.

\subsubsection{GHZ state in an amplitude damping (AD) channel}

The GHZ (Greenberger\textendash{}Horne\textendash{}Zeilinger) state
is a three qubit quantum state $\begin{array}{lcl}
\left|GHZ\right\rangle  & = & \frac{1}{\sqrt{2}}\left(\left|000\right\rangle +\left|111\right\rangle \right).\end{array}$ The first qubit of the state is acted upon by an amplitude damping
(AD) channel while the remaining two qubits remain unaffected. The
Kraus operators for AD channel for a single qubit state are as in
Eq. (\ref{eq:AD-kraussoperators}). Here, assuming that the three
qubits are independent of each other and do not have any interactions,
the Kraus operators for the action of AD channel only on the first
qubit, can be modeled as 
\begin{equation}
\begin{array}{lcl}
K_{1} & = & E_{0}\left(A\right)\otimes\mathbb{I}\left(B\right)\otimes\mathbb{I}\left(C\right),\\
K_{2} & = & E_{1}\left(A\right)\otimes\mathbb{I}\left(B\right)\otimes\mathbb{I}\left(C\right),
\end{array}\label{eq:KrausOp-GHZ_W}
\end{equation}
where $E_{0}$ and $E_{1}$ are as in Eq. (\ref{eq:AD-kraussoperators})
and $\mathbb{I}$ is a $2\times2$ identity matrix. Also, $A$, $B$
and $C$ stand for the first, second and third qubits of the GHZ state,
respectively. Thus, the density matrix of the GHZ state at time $t$
in the amplitude damping channel is 
\begin{equation}
\begin{array}{lcl}
\rho\left(t\right) & = & \overset{2}{\underset{i=1}{\Sigma}}K_{i}\left(t\right)\rho\left(0\right)K_{i}^{\dagger}\left(t\right),\end{array}\label{eq:density-mat-GHZ_W}
\end{equation}
where $\rho\left(0\right)=\left|GHZ\right\rangle \left\langle GHZ\right|,$
is the initial state at time $t=0$. Thus, at time $t$ the density
matrix for the GHZ state evolving in the presence of the AD channel
is 
\begin{equation}
\begin{array}{lcl}
\rho\left(t\right) & = & \frac{1}{2}\left[\begin{array}{cccccccc}
\left(1-\lambda\right) & 0 & 0 & 0 & 0 & 0 & 0 & \sqrt{\left(1-\lambda\right)}\\
0 & 0 & 0 & 0 & 0 & 0 & 0 & 0\\
0 & 0 & 0 & 0 & 0 & 0 & 0 & 0\\
0 & 0 & 0 & 0 & 0 & 0 & 0 & 0\\
0 & 0 & 0 & 0 & \lambda & 0 & 0 & 0\\
0 & 0 & 0 & 0 & 0 & 0 & 0 & 0\\
0 & 0 & 0 & 0 & 0 & 0 & 0 & 0\\
\sqrt{\left(1-\lambda\right)} & 0 & 0 & 0 & 0 & 0 & 0 & 1
\end{array}\right].\end{array}\label{eq:rho-AD-GHZ}
\end{equation}
Analytical expressions can be obtained for the different QDs for the
time evolved GHZ state described by Eq. (\ref{eq:rho-AD-GHZ}).

\begin{figure}
\centering{}\includegraphics[angle=-90,scale=0.6]{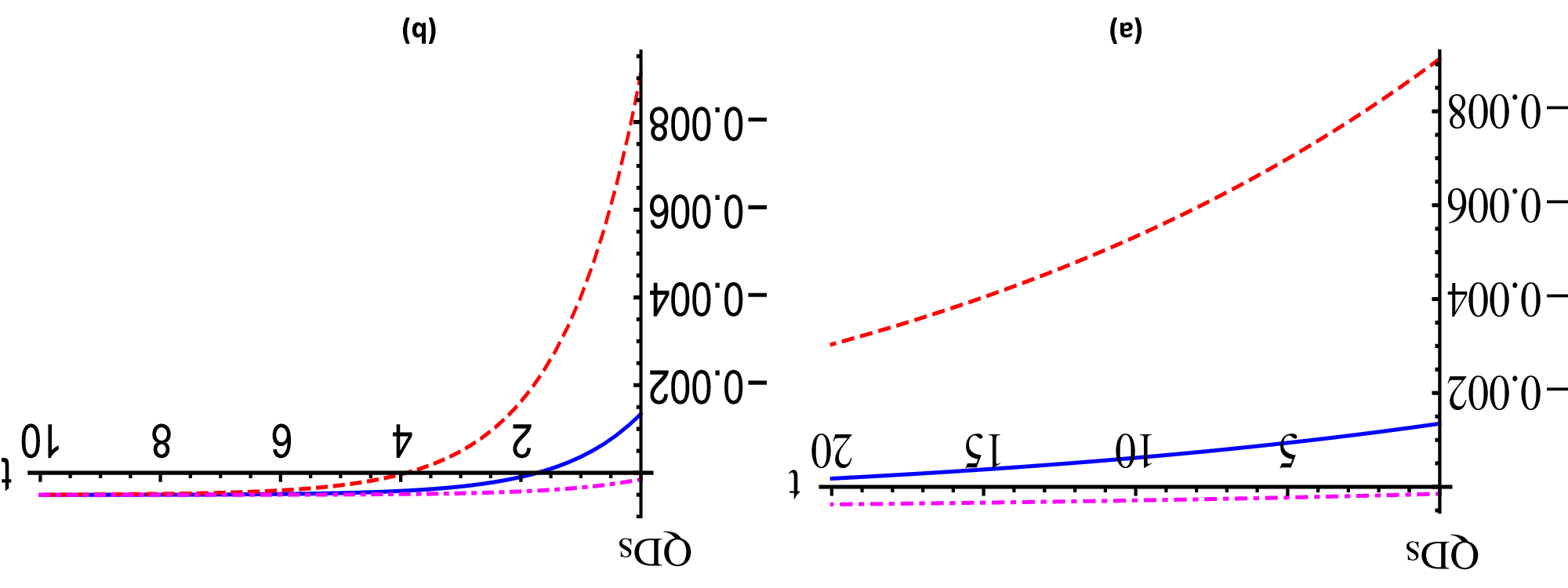}\caption{\label{fig:GHZ-AD}(Color online) 
The variation of all QDs with
time for the GHZ state when acted upon by (a) an AD noise on the first
qubit and (b) a generalized amplitude damping noise on each qubit, with $\gamma_{0}=0.1$ and 
$\theta_{1}=\frac{\pi}{2},\,\theta_{2}=\frac{\pi}{2},\,\theta_{3}=\frac{\pi}{2},\,\phi_{1}=\frac{\pi}{4},\,\phi_{2}=
\frac{\pi}{3},\,\phi_{3}=\frac{\pi}{6}$. In (b) different QDs are shown for $\omega=1.0$ 
with the first, second and third qubits subjected to generalized amplitude damping noise at $T=0$, 1 and 2, respectively. 
In both the plots smooth (blue), dashed (red) and dot-dashed (magenta) lines are for
the $W$, $P$ and $Q$ functions, respectively.}
\end{figure}

Specifically, we obtain the $W$ function as 
\begin{equation}
\begin{array}{lcl}
W\left(\theta_{1},\phi_{1},\theta_{2},\phi_{2},\theta_{3},\phi_{3}\right) & = & \frac{1}{64\pi^{3}}\left[1-\sqrt{3}\lambda\cos\theta_{1}+3\cos\theta_{2}\cos\theta_{3}+3\left(1-\lambda\right)\cos\theta_{1}\left(\cos\theta_{2}+\cos\theta_{3}\right)\right.\\
 & - & \left.3\sqrt{3}\left\{ \lambda\cos\theta_{1}\cos\theta_{2}\cos\theta_{3}-\sqrt{\left(1-\lambda\right)}\sin\theta_{1}\sin\theta_{2}\sin\theta_{3}\cos\left(\phi_{1}+\phi_{2}+\phi_{3}\right)\right\} \right],
\end{array}\label{eq:Wigner-AD-GHZ}
\end{equation}
while the $P$ function is 
\begin{equation}
\begin{array}{lcl}
P\left(\theta_{1},\phi_{1},\theta_{2},\phi_{2},\theta_{3},\phi_{3}\right) & = & \frac{1}{64\pi^{3}}\left[1+3\lambda\cos\theta_{1}+9\cos\theta_{2}\cos\theta_{3}+9\left(1-\lambda\right)\cos\theta_{1}\left(\cos\theta_{2}+\cos\theta_{3}\right)\right.\\
 & + & \left.27\left\{ \lambda\cos\theta_{1}\cos\theta_{2}\cos\theta_{3}-\sqrt{\left(1-\lambda\right)}\sin\theta_{1}\sin\theta_{2}\sin\theta_{3}\cos\left(\phi_{1}+\phi_{2}+\phi_{3}\right)\right\} \right],
\end{array}\label{eq:P-AD-GHZ}
\end{equation}
and the $Q$ function is 
\begin{equation}
\begin{array}{lcl}
Q\left(\theta_{1},\phi_{1},\theta_{2},\phi_{2},\theta_{3},\phi_{3}\right) & = & \frac{1}{64\pi^{3}}\left[1+\lambda\cos\theta_{1}+\cos\theta_{2}\cos\theta_{3}+\left(1-\lambda\right)\cos\theta_{1}\left(\cos\theta_{2}+\cos\theta_{3}\right)\right.\\
 & + & \left.\lambda\cos\theta_{1}\cos\theta_{2}\cos\theta_{3}-\sqrt{\left(1-\lambda\right)}\sin\theta_{1}\sin\theta_{2}\sin\theta_{3}\cos\left(\phi_{1}+\phi_{2}+\phi_{3}\right)\right].
\end{array}\label{eq:Q-AD-GHZ}
\end{equation}
The variation of the different QDs, as in Eqs. (\ref{eq:Wigner-AD-GHZ})-(\ref{eq:Q-AD-GHZ}), with 
time is depicted, for a particular
choice of the parameters, in Fig. \ref{fig:GHZ-AD}a. The
$W$ and $P$ functions exhibit negative values (nonclassical character)
for the times shown, which could be attributed to the initial entanglement
in the state. Further, for the sake of generality, we depict in Fig. \ref{fig:GHZ-AD}b the scenario wherein
all three qubits are affected by the generalized amplitude damping noise \cite{SGAD}, which can be obtained from 
Eq. (\ref{eq:SGAD-kraussoperators2-1}) by setting the bath squeezing parameters to zero. 
All three qubits are subjected to different 
temperatures corresponding to independent environment for each qubit, and simulates the scenario where 
each qubit travels through an independent channel. It can be observed that the nonclassicality indicated by the 
negativity of $W$ and $P$ functions at $t=0$ decays more rapidly when the last two qubits are subjected to 
finite temperature noises.

\subsubsection{W state in an amplitude damping (AD) channel}

\begin{figure}
\centering{}\includegraphics[angle=-90,scale=0.6]{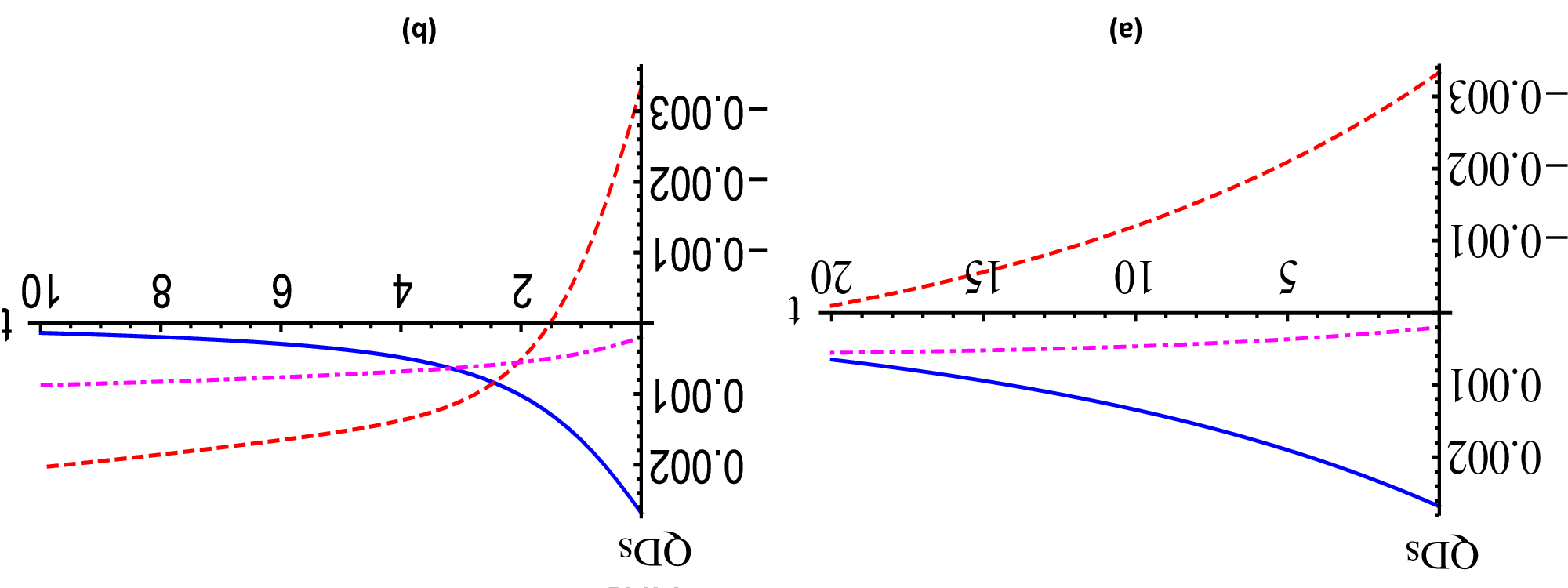}\caption{\label{fig:W-AD}(Color online) 
(a) Variation of  QDs with time for the
W state in the presence of an AD channel, acting only upon the first
qubit, with $\gamma_{0}=0.1$ and $\theta_{1}=\frac{\pi}{4},
\,\theta_{2}=\frac{\pi}{6},\,\theta_{3}=\frac{\pi}{3},\,\phi_{1}=\frac{\pi}{8},\,\phi_{2}=\frac{\pi}{4},
\,\phi_{3}=\frac{\pi}{6}$. (b) QDs for W state, when all the qubits are subjected to 
generalized amplitude damping noise, with $\omega=1.0$ and $T=0$, 1 and 2 for the first, second and third qubits, 
respectively. The remaining values of parameters in (b) are as in (a).
Smooth (blue), dashed (red) and dot-dashed (magenta) lines correspond to
$W$ function, $P$ function and $Q$ function, respectively.}
\end{figure}

For our second example of QDs of three qubit states, we take up the
W state $\begin{array}{lcl}
\left|W\right\rangle  & = & \frac{1}{\sqrt{3}}\left(\left|001\right\rangle +\left|010\right\rangle +\left|100\right\rangle \right).\end{array}$ As before, we consider the evolution where only the first qubit of
the state is acted upon by an AD channel. The Kraus operators, describing
the evolution are given by Eq. (\ref{eq:KrausOp-GHZ_W}). Here, as
in the case of GHZ or EPR states, assuming that the three qubits are
independent of each other and do not have any interactions, the density
matrix of the evolved state is, as in the last case, given by Eq.
(\ref{eq:density-mat-GHZ_W}), where $\rho\left(0\right)=\left|W\right\rangle \left\langle W\right|,$
is the initial state at time $t=0$. Hence, at time $t$, the W state
evolves, in the presence of the AD channel, to 
\begin{equation}
\begin{array}{lcl}
\rho\left(t\right) & = & \frac{1}{3}\left[\begin{array}{cccccccc}
0 & 0 & 0 & 0 & 0 & 0 & 0 & 0\\
0 & \left(1-\lambda\right) & \left(1-\lambda\right) & 0 & \sqrt{\left(1-\lambda\right)} & 0 & 0 & 0\\
0 & \left(1-\lambda\right) & \left(1-\lambda\right) & 0 & \sqrt{\left(1-\lambda\right)} & 0 & 0 & 0\\
0 & 0 & 0 & 0 & 0 & 0 & 0 & 0\\
0 & \sqrt{\left(1-\lambda\right)} & \sqrt{\left(1-\lambda\right)} & 0 & 1 & 0 & 0 & 0\\
0 & 0 & 0 & 0 & 0 & \lambda & \lambda & 0\\
0 & 0 & 0 & 0 & 0 & \lambda & \lambda & 0\\
0 & 0 & 0 & 0 & 0 & 0 & 0 & 0
\end{array}\right].\end{array}\label{eq:rho-AD-W}
\end{equation}
In this case, again, making use of Eq. (\ref{eq:rho-AD-W}), analytical
forms of the different QDs can be obtained as follows 
\begin{equation}
\begin{array}{lcl}
W\left(\theta_{1},\phi_{1},\theta_{2},\phi_{2},\theta_{3},\phi_{3}\right) & = & \frac{1}{64\pi^{3}}\left[1-\left(\cos\theta_{1}\cos\theta_{2}+\cos\theta_{2}\cos\theta_{3}+\cos\theta_{1}\cos\theta_{3}\right)+\frac{\sqrt{3}}{3}\left(\cos\theta_{1}+\cos\theta_{2}+\cos\theta_{3}\right)\right.\\
 & - & 3\sqrt{3}\cos\theta_{1}\cos\theta_{2}\cos\theta_{3}+2\left(1+\sqrt{3}\cos\theta_{1}\right)\sin\theta_{2}\sin\theta_{3}\cos\left(\phi_{2}-\phi_{3}\right)\\
 & + & 2\sqrt{\left(1-\lambda\right)}\left\{ \left(1+\sqrt{3}\cos\theta_{2}\right)\sin\theta_{1}\sin\theta_{3}\cos\left(\phi_{1}-\phi_{3}\right)\right.\\
 & + & \left.\left(1+\sqrt{3}\cos\theta_{3}\right)\sin\theta_{1}\sin\theta_{2}\cos\left(\phi_{1}-\phi_{2}\right)\right\} \\
 & + & \left.4\sqrt{3}\lambda\cos\theta_{1}\left\{ -\frac{1}{3}+\cos\theta_{2}\cos\theta_{3}-\sin\theta_{2}\sin\theta_{3}\cos\left(\phi_{2}-\phi_{3}\right)\right\} \right],
\end{array}\label{eq:Wigner-AD-W}
\end{equation}

\begin{equation}
\begin{array}{lcl}
P\left(\theta_{1},\phi_{1},\theta_{2},\phi_{2},\theta_{3},\phi_{3}\right) & = & \frac{1}{64\pi^{3}}\left[1-3\left(\cos\theta_{1}\cos\theta_{2}+\cos\theta_{2}\cos\theta_{3}+\cos\theta_{1}\cos\theta_{3}\right)-\left(\cos\theta_{1}+\cos\theta_{2}+\cos\theta_{3}\right)\right.\\
 & + & 27\cos\theta_{1}\cos\theta_{2}\cos\theta_{3}+6\left(1-3\cos\theta_{1}\right)\sin\theta_{2}\sin\theta_{3}\cos\left(\phi_{2}-\phi_{3}\right)\\
 & + & 6\sqrt{\left(1-\lambda\right)}\left\{ \left(1-3\cos\theta_{2}\right)\sin\theta_{1}\sin\theta_{3}\cos\left(\phi_{1}-\phi_{3}\right)\right.\\
 & + & \left.\left(1-3\cos\theta_{3}\right)\sin\theta_{1}\sin\theta_{2}\cos\left(\phi_{1}-\phi_{2}\right)\right\} \\
 & + & \left.4\lambda\cos\theta_{1}\left\{ 1-9\cos\theta_{2}\cos\theta_{3}+9\sin\theta_{2}\sin\theta_{3}\cos\left(\phi_{2}-\phi_{3}\right)\right\} \right],
\end{array}\label{eq:P-AD-W}
\end{equation}
and 
\begin{equation}
\begin{array}{lcl}
Q\left(\theta_{1},\phi_{1},\theta_{2},\phi_{2},\theta_{3},\phi_{3}\right) & = & \frac{1}{192\pi^{3}}\left[3-\left(\cos\theta_{1}\cos\theta_{2}+\cos\theta_{2}\cos\theta_{3}+\cos\theta_{1}\cos\theta_{3}\right)-\left(\cos\theta_{1}+\cos\theta_{2}+\cos\theta_{3}\right)\right.\\
 & + & 3\cos\theta_{1}\cos\theta_{2}\cos\theta_{3}+4\sin\frac{\theta_{1}^{2}}{2}\sin\theta_{2}\sin\theta_{3}\cos\left(\phi_{2}-\phi_{3}\right)\\
 & + & 4\sqrt{\left(1-\lambda\right)}\left\{ \sin\theta_{1}\sin\frac{\theta_{2}^{2}}{2}\sin\theta_{3}\cos\left(\phi_{1}-\phi_{3}\right)+\sin\theta_{1}\sin\theta_{2}\sin\frac{\theta_{3}^{2}}{2}\cos\left(\phi_{1}-\phi_{2}\right)\right\} \\
 & + & \left.4\lambda\cos\theta_{1}\left\{ 1-\cos\theta_{2}\cos\theta_{3}+\sin\theta_{2}\sin\theta_{3}\cos\left(\phi_{2}-\phi_{3}\right)\right\} \right].
\end{array}\label{eq:Q-AD-W}
\end{equation}
The variation of the QDs with time is shown in Fig. \ref{fig:W-AD}a
for a particular choice of the parameters. Here, the $P$ function
exhibits negative values for the times shown, but in contrast to the
GHZ case, the $W$ function is found to be positive. Thus, the signature
of quantumness (nonclassicality) of the state is identified by $P$
function, but $W$ function fails to detect the same. Similar to the QDs of GHZ state in the presence
of generalized amplitude damping noise, a decrease in the nonclassicality of W state with time can be observed due to the 
effect of noise on the last two qubits at non-zero temperatures (cf. Fig. \ref{fig:W-AD}b).

\subsection{$N$ qubit Dicke Model}

We conclude our discussion of QDs for two-level systems (qubits) with
the Dicke model. The interaction of a collection of identical two-level
atoms with a single mode of quantized field is the Dicke model \citep{dicke,tacum}
and is the multi-atom generalization of the Jaynes-Cummings model.
The Hamiltonian for the Dicke model is given as 
\begin{equation}
H=\omega\left(a^{\dagger}a+\Sigma_{z}^{c}\right)+g\left(a\Sigma_{+}^{c}+a^{\dagger}\Sigma_{-}^{c}\right).\label{eq:Dicke-model}
\end{equation}
Here, $\omega$ is the resonant atomic frequency, $g$ is the coupling
constant, $a$ ($a^{\dagger}$) is the annihilation (creation) operator
for the radiation field, and $\Sigma_{\pm}^{c}=\overset{N}{\underset{j=1}{\sum}}{\sigma^{\left(j\right)}}_{\pm},$
and $\Sigma_{z}^{c}=\frac{1}{2}\overset{N}{\underset{j=1}{\sum}}{\sigma^{\left(j\right)}}_{z},$
where $N$ is the number of atoms considered. Also, the symbol $c$
in the superscript stands for collective. A generalization of this
model, dynamics of a collection of atoms interacting with a squeezed
radiation field, was made in \citep{AgPu}. The Dicke model has analytical
solutions in two regimes: weak and strong field regimes, which corresponds
to average photon number being much smaller or greater than the number
of atoms, respectively.

\begin{figure}
\centering{}\includegraphics[scale=0.6]{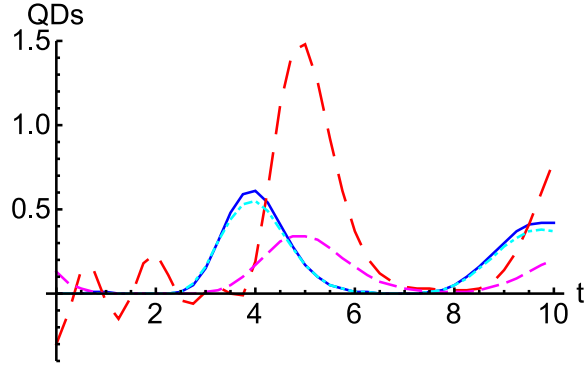}\caption{\label{fig:Dicke}(Color online) Variation of all QDs with time for
the $N=4$ atoms Dicke model with $\theta=\frac{\pi}{3},$ and $\phi=\frac{\pi}{2}.$
Smooth (blue), large dashed (red), small dashed (magenta) and dot-dashed
(cyan) lines are for $W$ function, $P$ function, $Q$ function,
and $F$ function, respectively. The four atom case is equivalent
to the spin-2 case and consequently different values of the $W$ and
$F$ functions can be observed.}
\end{figure}

We discuss, here, only the second case in the presence of strong initial
field \citep{DM}, i.e., $\bar{n}\gg N$, where $\bar{n}$ is the
average number of photons in the initial coherent field. In the dissipative
case, modeling the microwave region of zero temperature dissipative
cavity quantum electrodynamics, restricting ourselves to the condition
$\frac{\gamma}{g}\ll\sqrt{\bar{n}}$, the master equation in Fock
basis is 
\begin{equation}
\begin{array}{lcl}
\dot{\rho}_{nm} & = & -2ig\left(\Sigma_{x}^{c}\sqrt{n_{N}}\rho_{nm}-\rho_{nm}\sqrt{m_{N}}\Sigma_{x}^{c}\right)+\frac{\gamma}{2}\left(2a\rho a^{\dagger}-a^{\dagger}a\rho-\rho a^{\dagger}a\right)_{nm},\end{array}\label{eq:master-eq-DM}
\end{equation}
where $n_{N}=n+\frac{1}{2}-\frac{N}{2}$, and $\gamma$ is the cavity
decay constant. This master equation takes a simple form in the dressed
atomic basis, in which the atomic operator $\Sigma_{x}^{c}=\frac{1}{2}(\Sigma_{+}^{c}+\Sigma_{-}^{c})$
is diagonal 
\[
\Sigma_{x}^{c}\left|\tilde{q}\right\rangle =\lambda_{q}\left|\tilde{q}\right\rangle ,
\]
where $\lambda_{q}=q-\frac{N}{2},$ and $q=0,1,\ldots,N.$ These dressed
atomic states can be expressed in terms of the bare atomic basis as
$\left|\tilde{q}\right\rangle =\underset{q}{\sum}C_{qk}\left|k\right\rangle ,$
where $C_{qk}=\left\langle k\left|\tilde{q}\right.\right\rangle =i^{q-k}d_{qk}^{N}\left(-\frac{\pi}{2}\right),$
and $d_{qk}^{N}\left(\theta\right)$ are Wigner $d$ functions. The
bare atomic basis (Dicke states) is defined as 
\[
\Sigma_{z}^{c}\left|k\right\rangle =\left(k-\frac{N}{2}\right)\left|k\right\rangle ,\,\,\,\,0\leq k\leq N,
\]
where $k$ is the number of excited atoms.

Now, we consider the initial atoms-field density matrix 
\[
\rho\left(0\right)=\left|{\rm atin}\right\rangle \left\langle {\rm atin}\right|\otimes\left|\alpha\right\rangle \left\langle \alpha\right|,
\]
where $\left|{\rm atin}\right\rangle $ is the initial atomic state,
taken here to be ground state, i.e., none of the atoms are excited
and $\left|\alpha\right\rangle $ denotes the initial strong coherent
state of the field. The atomic density matrix, in the bare atomic
basis, after tracing out the radiation field evolves as 
\begin{equation}
\begin{array}{lcl}
\rho_{kl}\left(t\right) & = & \underset{n}{\sum}\alpha_{k+n}\left(t\right)\alpha_{l+n}^{*}\left(t\right)\overset{N}{\underset{q,p=0}{\Sigma}}C_{kq}C_{lp}C_{0q}C_{0p}^{*}G_{pq}^{n+k,n+l}\left(t\right),\end{array}\label{eq:density-mat-t}
\end{equation}
where 
\[
G_{qp}^{n,m}\left(t\right)=\exp\left[-2igt\left(\lambda_{q}\sqrt{n_{N}}-\lambda_{p}\sqrt{m_{N}}\right)-\Theta_{qp}\left(t\right)\right],
\]
and $\Theta_{qp}\left(t\right)=\bar{n}\left\{ \left(1-e^{-\gamma t}\right)-\frac{\gamma}{\gamma^{\prime}}\left(1-e^{-\gamma^{\prime}t}\right)\right\} $,
$\gamma^{\prime}=\gamma+\frac{ig\left(\lambda_{q}-\lambda_{p}\right)}{\sqrt{\bar{n}\left(t\right)-\frac{N}{2}+\frac{1}{2}}}$
and $\alpha_{n}\left(t\right)=e^{-\bar{n}\left(t\right)/2}\frac{\bar{n}\left(t\right)^{n/2}}{\sqrt{n!}},$
with $\bar{n}\left(t\right)=\bar{n}e^{-\gamma t}.$

The evolution of all the QDs can be obtained for the Dicke model in
the dissipative case using Eq. (\ref{eq:density-mat-t}). Here, we
consider the $N=4$ atoms Dicke model in a strong initial field with
average photon number 30. Further, we take $\frac{\gamma}{g}=10^{-2},$
and coupling constant as 0.1. To achieve consistence with the notations
used in this article, we have used $j=\frac{N}{2}$, and $m,\, m^{\prime}=-j,\ldots,j$.
The multipole operators required to calculate the QDs are equivalent
to spin-2 multipole operators and can be seen in Appendix. In Fig.
\ref{fig:Dicke}, variation of the various QDs, as they evolve with
time, is depicted. The $P$ function exhibits negative values at some
early times, indicative of the quintessence of quantumness in the
system, but eventually becomes positive due to dissipative effects.
Different values of the $W$ and $F$ QDs are observed at some time
intervals, consistent with the observation that the four atom case
is equivalent to the spin-2 case. For this specific choice of parameters
and restricting ourselves to the computational accuracy of the numerical
method adopted here, we did not observe any signature of nonclassicality
via $W$ function. However, $W$ function can witness the nonclassical
characteristics present in the Dicke model for other values of $\theta$
and $\phi$. This point will be clearly illustrated in Fig. \ref{fig:W_Vol}
d, where we show positive values of nonclassical volume (which is
obtained by integrating the modulus of $W$ function over all possible
values of $\theta$ and $\phi$) for the Dicke model. These positive
values of nonclassical volume imply negative values of $W$ function
for some values of $\theta$ and $\phi$ (cf. (Eq. \ref{eq:ncv})).

\section{{\normalsize{{QDs for a spin-1 state}}}}

\begin{figure}
\begin{centering}
\includegraphics[angle=-90,scale=0.5]{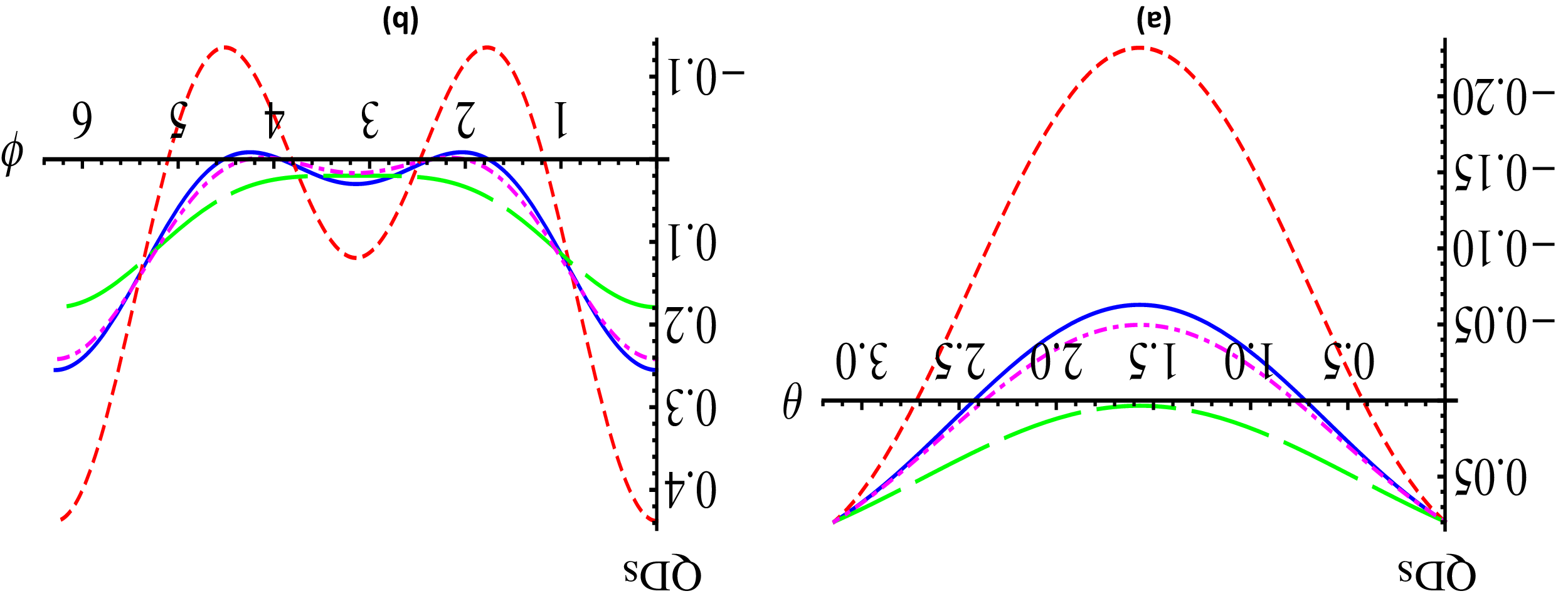} 
\par\end{centering}

\caption{\label{fig:spin-1-2D}(Color online) Variation of all the QDs is shown
for the spin-1 state with $a_{+}=a_{0}=a_{-}=\frac{1}{\sqrt{3}}.$
(a) shows $W$ (smooth blue line), $P$ (small dashed red line), $Q$
(large dashed green line), and $F$ (dot-dashed magenta line) functions
with $\phi=\frac{2\pi}{3}.$ Similarly, (b) shows all the QDs for
$\theta=\frac{\pi}{4}.$ Both (a) and (b) clearly bring out the point
that $W$ and $F$ functions are not equal, in general, in contrast
to the spin-$\frac{1}{2}$ cases.}
\end{figure}

Now, we extend the discussion of spin QDs from spin-$\frac{1}{2}$
to spin-1 states. For a spin-1 pure state \citep{blum} 
\[
\left|\chi\right\rangle =a_{+}\left|+1\right\rangle +a_{0}\left|0\right\rangle +a_{-}\left|-1\right\rangle ,
\]
QDs can be constructed using appropriate multipole operators and spherical
harmonics. A few relevant multipole operators are $T_{00}=\frac{1}{\sqrt{3}}\left[\begin{array}{ccc}
1 & 0 & 0\\
0 & 1 & 0\\
0 & 0 & 1
\end{array}\right],$ $T_{11}=\frac{1}{\sqrt{2}}\left[\begin{array}{ccc}
0 & -1 & 0\\
0 & 0 & -1\\
0 & 0 & 0
\end{array}\right],$ $T_{10}=\frac{1}{\sqrt{2}}\left[\begin{array}{ccc}
1 & 0 & 0\\
0 & 0 & 0\\
0 & 0 & -1
\end{array}\right],$ $T_{22}=\left[\begin{array}{ccc}
0 & 0 & 1\\
0 & 0 & 0\\
0 & 0 & 0
\end{array}\right],$ $T_{21}=\frac{1}{\sqrt{2}}\left[\begin{array}{ccc}
0 & -1 & 0\\
0 & 0 & 1\\
0 & 0 & 0
\end{array}\right],$ and $T_{20}=\frac{1}{\sqrt{6}}\left[\begin{array}{ccc}
1 & 0 & 0\\
0 & -2 & 0\\
0 & 0 & 1
\end{array}\right].$ All other multipole operators can be obtained from these operators.
The analytical expressions of the different QDs are obtained as 
\begin{equation}
\begin{array}{lcl}
W\left(\theta,\phi\right) & = & \frac{1}{16\pi}\left[4-\sqrt{10}+3\sqrt{10}\left(\cos^{2}\theta+\left|a_{0}\right|^{2}-3\left|a_{0}\right|^{2}\cos^{2}\theta\right)+6\sqrt{2}\left(\left|a_{+}\right|^{2}-\left|a_{-}\right|^{2}\right)\cos\theta\right.\\
 & + & \left.\left(6a_{0}\sin\theta\left\{ a_{+}^{*}\exp(-i\phi)\left(1+\sqrt{5}\cos\theta\right)+a_{-}^{*}\exp(i\phi)\left(1-\sqrt{5}\cos\theta\right)\right\} +3\sqrt{10}a_{+}a_{-}^{*}\sin^{2}\theta\exp(2i\phi)+{\rm c.c.}\right)\right],
\end{array}\label{eq:W-spin1}
\end{equation}
\begin{equation}
\begin{array}{lcl}
P\left(\theta,\phi\right) & = & \frac{3}{8\pi}\left[-1+5\left(\cos^{2}\theta+\left|a_{0}\right|^{2}-3\left|a_{0}\right|^{2}\cos^{2}\theta\right)-2\left(\left|a_{+}\right|^{2}-\left|a_{-}\right|^{2}\right)\cos\theta\right.\\
 & + & \left.\left(\sqrt{2}a_{0}\sin\theta\left\{ a_{+}^{*}\exp(-i\phi)\left(1-5\cos\theta\right)+a_{-}^{*}\exp(i\phi)\left(1+5\cos\theta\right)\right\} +5a_{+}a_{-}^{*}\sin^{2}\theta\exp(2i\phi)+{\rm c.c.}\right)\right],
\end{array}\label{eq:P-spin1}
\end{equation}
\begin{equation}
\begin{array}{lcl}
Q\left(\theta,\phi\right) & = & \frac{3}{16\pi}\left[1+\left(\cos^{2}\theta+\left|a_{0}\right|^{2}-3\left|a_{0}\right|^{2}\cos^{2}\theta\right)-2\left(\left|a_{+}\right|^{2}-\left|a_{-}\right|^{2}\right)\cos\theta\right.\\
 & + & \left.\left(\sqrt{2}a_{0}\sin\theta\left\{ a_{+}^{*}\exp(-i\phi)\left(1-\cos\theta\right)+a_{-}^{*}\exp(i\phi)\left(1+\cos\theta\right)\right\} +5a_{+}a_{-}^{*}\sin^{2}\theta\exp(2i\phi)+{\rm c.c.}\right)\right],
\end{array}\label{eq:Q-spin1}
\end{equation}
and 
\begin{equation}
\begin{array}{lcl}
F\left(\theta,\phi\right) & = & \frac{3}{32\pi}\left[1+5\left(\cos^{2}\theta+\left|a_{0}\right|^{2}-3\left|a_{0}\right|^{2}\cos^{2}\theta\right)+4\sqrt{2}\left(\left|a_{+}\right|^{2}-\left|a_{-}\right|^{2}\right)\cos\theta\right.\\
 & + & \left.\left(a_{0}\sin\theta\left\{ a_{+}^{*}\exp(-i\phi)\left(4+5\sqrt{2}\cos\theta\right)+a_{-}^{*}\exp(i\phi)\left(4-5\sqrt{2}\cos\theta\right)\right\} +5a_{+}a_{-}^{*}\sin^{2}\theta\exp(2i\phi)+{\rm c.c.}\right)\right].
\end{array}\label{eq:F-spin1}
\end{equation}
In Fig. \ref{fig:spin-1-2D}, we illustrate the behavior of the above
QDs for $a_{+}=a_{0}=a_{-}=\frac{1}{\sqrt{3}}$. Here, we do not consider
the effect of noise on the evolution of the QDs, a topic to which
we will return back to, in the future. The purpose here, besides studying
the quantumness in the system via the QDs, is to emphasize the nonequivalence,
for the spin-1 case, of the $W$ and $F$ functions, in contrast to
the spin-$\frac{1}{2}$ case. Fig. \ref{fig:spin-1-2D} depicts the
behavior of the QDs with respect to $\theta$ and $\phi$, both of
which show a symmetric behavior about the central point on the ordinate.
The $P$, $W$ and $F$ functions exhibit negative values, indicative
of the quantumness in the system, with the $P$ function being the
most sensitive indicator, as expected. Also, the $F$ and $W$ functions
are clearly distinct.

\section{Nonclassical volume}

Till now, we have studied nonclassicality using negative values of
the $W$ or $P$ function. Negative values of the QDs only provide
a signature of nonclassicality, but they do not provide a quantitative
measure of nonclassicality. There do exist some quantitative measures
of nonclassicality, see for example, \citep{with-adam-meas} for a
review. One such measure is nonclassical volume introduced in \citep{zyco}.
In this approach, the doubled volume of the integrated negative part of
the $W$ function of a given quantum state is used as a quantitative measure of the quantumness \citep{zyco}. Using our knowledge of the $W$
functions for various systems, studied here, the nonclassical volume
$\delta$, which is defined as 
\begin{equation}
\delta=\int\left|W\left(\theta,\phi\right)\right|\sin\theta d\theta d\phi-1,\label{eq:ncv}
\end{equation}
can be computed. It can be easily observed that a nonzero value of
$\delta$ would imply the existence of nonclassicality, but this measure
is not useful in measuring inherent nonclassicality in all quantum
states. This is so because, the $W$ function is only a witness of
nonclassicality (it does not provide a necessary condition). However,
this measure of nonclassicality has been used in a number of optical
systems, see for example, \citep{with-adam,with-jay} and references
therein.

\begin{figure}
\begin{centering}
\includegraphics[angle=-90,scale=0.8]{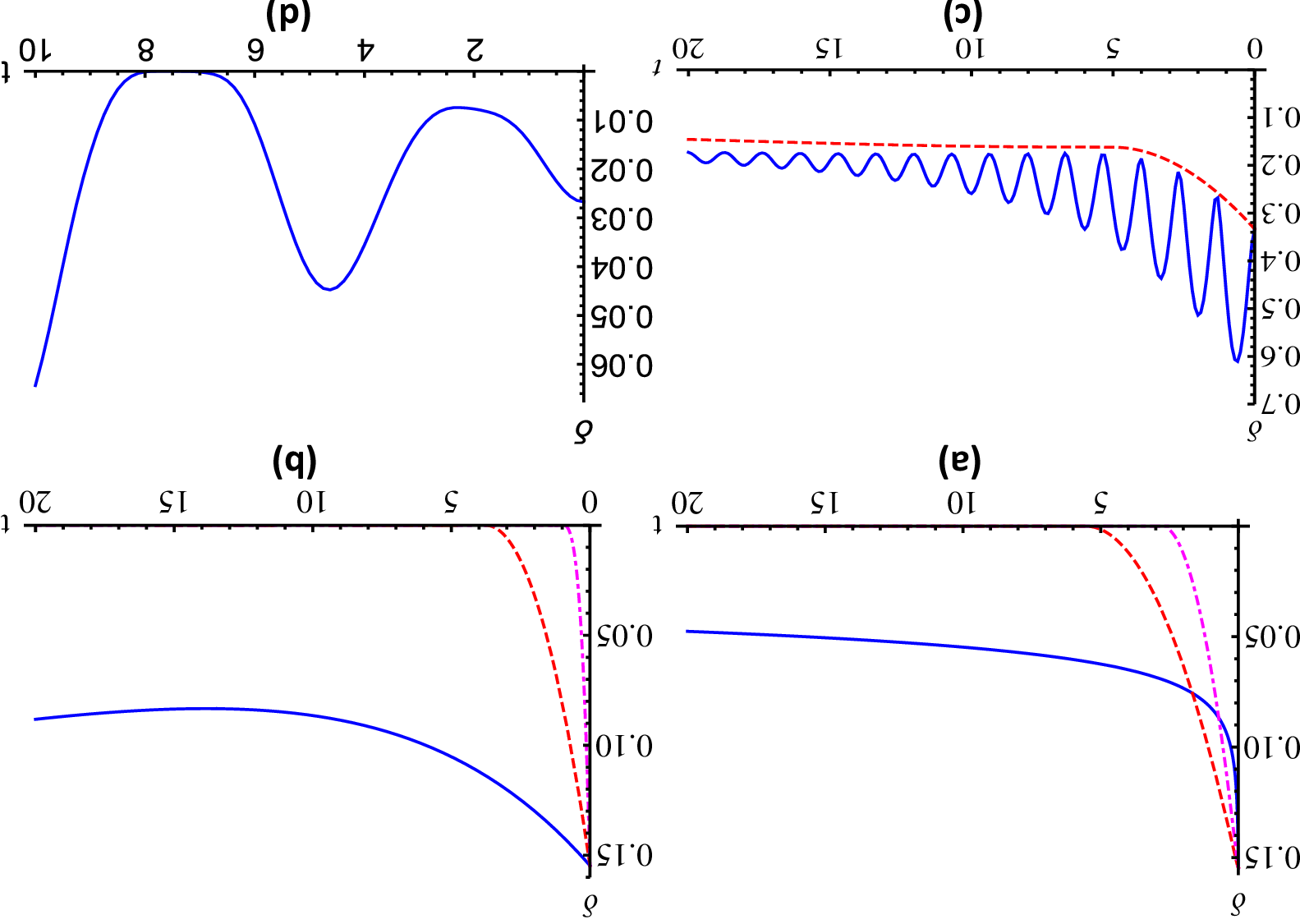} 
\par\end{centering}

\caption{\label{fig:W_Vol}(Color online) The plots (a)-(d) depict the variation
of the nonclassical volume in the presence of different noises. (a)
The variation of the nonclassical volume with time is shown for the
single spin-$\frac{1}{2}$ atomic coherent state in the presence of
the QND noise with $\gamma_{0}=0.1,\, r=0,\, a=0,\,\omega_{c}=100,\,\omega=1.0,$
and $\alpha=\frac{\pi}{2},\,\beta=\frac{\pi}{3},$ in the units of
$\hbar=k_{B}=1$, where the smooth (blue), dashed (red) and dot-dashed
(magenta) lines correspond to different temperatures $T=0,\,1$ and
$2$, respectively. (b) The variation of the nonclassical volume with
time is shown for a single spin-$\frac{1}{2}$ atomic coherent state
in a SGAD channel, where the smooth (blue) line corresponds to the
variation in nonclassical volume in vacuum bath, i.e., at $T=0$ and
squeezing parameters $r=\xi=0$ (amplitude damping channel); dashed
(red) line corresponds to the variation in a channel with zero squeezing
at $T=3$, i.e., generalized amplitude damping channel, and dot-dashed
(magenta) line corresponds to the variation with squeezing $r=1,$
squeezing angle $\xi=0,$ and $T=3$. In all these cases $\alpha$
and $\beta$ have the same values as in (a). (c) The behavior of nonclassical
volume with time is depicted in a vacuum bath for the state discussed
in Section \ref{sub:Vacuum-bath} with the inter-qubit spacing $r_{12}=0.05$
(smooth blue line), and $r_{12}=2.0$ (red dashed line). (d) The temporal
behavior of nonclassical volume of the four atom Dicke model interacting
with a strong input coherent field with average photon number 30,
in the dissipative regime with $\frac{\gamma}{g}=10^{-2},$ and coupling
constant as 0.1.}
\end{figure}

Here, we will illustrate the time evolution of $\delta$ for some
of the spin-qubit systems, studied above. Specifically, Fig. \ref{fig:W_Vol}
a and b show the variation of $\delta$ for two spin-$\frac{1}{2}$
systems, initially in an atomic coherent state under the influence
of QND and SGAD channels, respectively. Fig. \ref{fig:W_Vol} c shows
two spin-$\frac{1}{2}$ states in a two-qubit vacuum bath. The dashed
and dot-dashed lines in Fig. \ref{fig:W_Vol} a and b, i.e., for atomic
coherent state in QND with finite temperature and in SGAD with finite
temperature and squeezing, exhibit the exponential reduction of nonclassical
volume with time implying a quick transition from nonclassical to
classical states, whereas the smooth lines in a and b (when temperature
and squeezing parameters are taken to be zero) show that after an
initial reduction, the nonclassical volume stabilizes over a reasonably
large duration. Thus, nonclassicality does not get completely destroyed
with time. A similar nature of time evolution of $\delta$ is also
observed for the dashed line in Fig. \ref{fig:W_Vol} c (a two qubit
state in a vacuum bath with relatively large inter-qubit spacing),
whereas an oscillatory nature is observed for small inter-qubit spacing,
depicted here by a smooth line. It should be noted that the smooth blue line in Fig. \ref{fig:W_Vol} b 
corresponds to the nonclassical volume for an
atomic coherent state dissipatively interacting with a vacuum bath, i.e., at zero temperature and squeezing, 
while the $W$ function of the atomic coherent state, 
illustrated in Fig. \ref{fig:SGAD-ACS}, is for non-zero temperature and squeezing. At zero temperature, 
the nonclassicality present in the system is expected to 
survive for a relatively longer period of time. Further, the nonclassical volume is the overall contribution in nonclassicality
from all values of $\theta$ and $\phi$. It is possible that the $W$ function 
shown for a particular value of $\theta$ and $\phi$, in the previous sections, may not exibit nonclassical behaviour at 
time $t$ whereas the other possible values provide a finite contribution to nonclassical volume, 
resulting in a nonvanishing $\delta$. Interestingly, in Fig. \ref{fig:W_Vol} d for the nonclassical volume of the 
Dicke model, we find that the
amount of nonclassicality oscillates with time. This seems to arise from the weakness of the nonclassicality measure 
used here.

\section{Conclusions}

The nonclassical nature of all the systems studied here, of relevance
to the fields of quantum optics and information, is illustrated via
their quasiprobability distributions as a function of the time of
evolution as well as various state or bath parameters. We also provide
a quantitative idea of the amount of nonclassicality observed in some
of the systems studied using a measure which essentially makes use
of the $W$ function. These issues assume significance in questions
related to quantum state engineering, where the central point is to
have a clear understanding of coherences in the quantum mechanical
system being used. Thus, it is essential to have an understanding
over quantum to classical transitions, under ambient conditions. This
is made possible by the present work, where a comprehensive analysis
of QDs for spin-qubit systems is made under general open system effects,
including both pure dephasing as well as dissipation, making it relevant
from the perspective of experimental implementation. Along with the
well known $W$, $P$ and $Q$ quasiprobability distributions, we
also discuss the so called $F$ function and specify its relation
to the $W$ function. We expect this work to have an impact on issues
related to state reconstruction, in the presence of decoherence and
dissipation. These quasiprobability distributions also play an important
role in fundamental issues such as complementarity between number
and phase distributions as well as for phase dispersion in atomic
systems. It is interesting to note that in \citep{victor}, a connection
was established between negative values of a particular quasiprobability
and potential for quantum speed-up. The present study could be of
use to probe this connection deeper.

\textbf{Acknowledgement: }A. P. and K.T. thank Department of Science
and Technology (DST), India for support provided through the DST project
No. SR/S2/LOP-0012/2010. SB thanks R. Srikanth for some useful discussions
during the early stages of this work. We also thank Usha Devi for
a number of helpful comments during various stages of this work.

\section*{Appendix: Multipole operators for Dicke model}

Here, we collect the multipole operators used for the computation
of QDs for the four qubit Dicke model in Sec. (IV.D).

$T_{00}=\frac{1}{\sqrt{5}}\left[\begin{array}{ccccc}
1 & 0 & 0 & 0 & 0\\
0 & 1 & 0 & 0 & 0\\
0 & 0 & 1 & 0 & 0\\
0 & 0 & 0 & 1 & 0\\
0 & 0 & 0 & 0 & 1
\end{array}\right],$ $T_{11}=\frac{1}{\sqrt{10}}\left[\begin{array}{ccccc}
0 & 0 & 0 & 0 & 0\\
-\sqrt{2} & 0 & 0 & 0 & 0\\
0 & -\sqrt{3} & 0 & 0 & 0\\
0 & 0 & -\sqrt{3} & 0 & 0\\
0 & 0 & 0 & -\sqrt{2} & 0
\end{array}\right],$ $T_{10}=\frac{1}{\sqrt{10}}\left[\begin{array}{ccccc}
-2 & 0 & 0 & 0 & 0\\
0 & -1 & 0 & 0 & 0\\
0 & 0 & 0 & 0 & 0\\
0 & 0 & 0 & 1 & 0\\
0 & 0 & 0 & 0 & 2
\end{array}\right],$

$T_{22}=\frac{1}{\sqrt{7}}\left[\begin{array}{ccccc}
0 & 0 & 0 & 0 & 0\\
0 & 0 & 0 & 0 & 0\\
\sqrt{2} & 0 & 0 & 0 & 0\\
0 & \sqrt{3} & 0 & 0 & 0\\
0 & 0 & \sqrt{2} & 0 & 0
\end{array}\right],$ $T_{21}=\frac{1}{\sqrt{14}}\left[\begin{array}{ccccc}
0 & 0 & 0 & 0 & 0\\
\sqrt{6} & 0 & 0 & 0 & 0\\
0 & 1 & 0 & 0 & 0\\
0 & 0 & -1 & 0 & 0\\
0 & 0 & 0 & -\sqrt{6} & 0
\end{array}\right],$ $T_{20}=\frac{1}{\sqrt{14}}\left[\begin{array}{ccccc}
2 & 0 & 0 & 0 & 0\\
0 & -1 & 0 & 0 & 0\\
0 & 0 & -2 & 0 & 0\\
0 & 0 & 0 & 1 & 0\\
0 & 0 & 0 & 0 & 2
\end{array}\right],$

$T_{33}=-\frac{1}{\sqrt{2}}\left[\begin{array}{ccccc}
0 & 0 & 0 & 0 & 0\\
0 & 0 & 0 & 0 & 0\\
0 & 0 & 0 & 0 & 0\\
1 & 0 & 0 & 0 & 0\\
0 & 1 & 0 & 0 & 0
\end{array}\right],$ $T_{32}=\frac{1}{\sqrt{2}}\left[\begin{array}{ccccc}
0 & 0 & 0 & 0 & 0\\
0 & 0 & 0 & 0 & 0\\
-1 & 0 & 0 & 0 & 0\\
0 & 0 & 0 & 0 & 0\\
0 & 0 & 1 & 0 & 0
\end{array}\right],$ $T_{31}=\frac{1}{\sqrt{10}}\left[\begin{array}{ccccc}
0 & 0 & 0 & 0 & 0\\
-\sqrt{3} & 0 & 0 & 0 & 0\\
0 & \sqrt{2} & 0 & 0 & 0\\
0 & 0 & \sqrt{2} & 0 & 0\\
0 & 0 & 0 & -\sqrt{3} & 0
\end{array}\right],$

$T_{30}=\frac{1}{\sqrt{10}}\left[\begin{array}{ccccc}
-1 & 0 & 0 & 0 & 0\\
0 & 2 & 0 & 0 & 0\\
0 & 0 & 0 & 0 & 0\\
0 & 0 & 0 & -2 & 0\\
0 & 0 & 0 & 0 & 1
\end{array}\right],$ $T_{44}=\left[\begin{array}{ccccc}
0 & 0 & 0 & 0 & 0\\
0 & 0 & 0 & 0 & 0\\
0 & 0 & 0 & 0 & 0\\
0 & 0 & 0 & 0 & 0\\
1 & 0 & 0 & 0 & 0
\end{array}\right],$ $T_{43}=\frac{1}{\sqrt{2}}\left[\begin{array}{ccccc}
0 & 0 & 0 & 0 & 0\\
0 & 0 & 0 & 0 & 0\\
0 & 0 & 0 & 0 & 0\\
1 & 0 & 0 & 0 & 0\\
0 & -1 & 0 & 0 & 0
\end{array}\right],$

$T_{42}=\frac{1}{\sqrt{14}}\left[\begin{array}{ccccc}
0 & 0 & 0 & 0 & 0\\
0 & 0 & 0 & 0 & 0\\
\sqrt{3} & 0 & 0 & 0 & 0\\
0 & -2\sqrt{2} & 0 & 0 & 0\\
0 & 0 & \sqrt{3} & 0 & 0
\end{array}\right],$ $T_{41}=\frac{1}{\sqrt{14}}\left[\begin{array}{ccccc}
0 & 0 & 0 & 0 & 0\\
1 & 0 & 0 & 0 & 0\\
0 & -\sqrt{6} & 0 & 0 & 0\\
0 & 0 & \sqrt{6} & 0 & 0\\
0 & 0 & 0 & -1 & 0
\end{array}\right],$ and $T_{40}=\frac{1}{\sqrt{70}}\left[\begin{array}{ccccc}
1 & 0 & 0 & 0 & 0\\
0 & -4 & 0 & 0 & 0\\
0 & 0 & 6 & 0 & 0\\
0 & 0 & 0 & -4 & 0\\
0 & 0 & 0 & 0 & 1
\end{array}\right].$

\end{document}